# Labor Informality and Credit Market Accessibility

Alina Malkova
UNC Chapel Hill

Klara Sabirianova Peter
UNC Chapel Hill

Jan Svejnar
Columbia University,
CERGE-EI, CEPR and IZA

The paper investigates the effects of the credit market development on the labor mobility between the informal and formal labor sectors. In the case of Russia, due to the absence of a credit score system, a formal lender may set a credit limit based on the verified amount of income. To get a loan, an informal worker must first formalize his or her income (switch to a formal job), and then apply for a loan. To show this mechanism, the RLMS data was utilized, and the empirical method is the dynamic multinomial logit model of employment. The empirical results show that a relaxation of credit constraints increases the probability of transition from an informal to a formal job, and improved CMA (by one standard deviation) increases the chances of informal sector workers to formalize by 5.4 ppt. These results are robust in different specifications of the model. Policy simulations show strong support for a reduction in informal employment in response to better CMA in credit-constrained communities.

# 1. Introduction

In this paper, we provide a simple theoretical model and an in-depth empirical estimation of the link between credit market development and labor mobility among three labor market states: informal sector, formal sector and non-employment. There are large literatures related, respectively, to informality, credit markets and labor mobility. However, our paper is motivated by the fact that the literatures have not explored the effects of credit market development on labor mobility between the informal and formal sectors.

One strand of the literature examines the informal sector. Djankov, Lieberman, Mukherjee and Nenova (2002) for instance explore the benefits and costs of informality, noting that informal activities are costly to entrepreneurs who operate them because they cannot utilize government and some private sector services available to firms that fully comply with regulations. However, many transition economies have experienced a surge in informal business activities because the quality of public services is poor and there are few tangible benefits to going formal. The authors note that benefits of informality lie in the avoidance of taxes, fees for licenses, permits and other regulatory charges, while the costs of informality are the implicit tax in the form of bribery, lack of safety net such as insurance and pension, and the inability to tap into formal credit markets. The latter observation is a key motivation for our study.

Gokalp, Lee and Peng (2017) in turn note that the degree of competition from the informal economy may affect the decision of formal firms to not fully comply with regulatory authorities. Informal firms may have inherent advantages through cost savings by not complying with government regulations. Using data from the World Bank's Productivity and Investment Climate Survey, the authors find that when the cost of compliance is high, firms will lower their compliance levels to stay competitive with the informal sector. On the other hand, when institutions provide sufficient advantages to firms staying in the formal sector, such as through easier access to credit, firms are more likely to comply with regulations and stay in the formal sector. The study is relevant for our analysis in that it suggests that credit provision may have an effect on the formality v. informality decision of firms.

Another large literature focuses on the part played by formal v. informal credit. Nguimkeu (2014) is an example of studies that focus on the provision of credit to informal sector entrepreneurs. Using data from a cross-sectional sample of Cameroon households in the National Survey on Employment and Informal Sector (EESI), the author finds microfinance programs that provide credit access to the poor reduce the size of subsistence economy by up to 10% while doubling the share of entrepreneurs in the informal sectors. Microfinance can also improve the total earnings capacity of the informal sector by up to 30% and has the potential of lowering heterogeneity by reducing the misallocation of skills and capital. In another study, using data from the Ghana Living Standards Survey, Akudugu (2014) finds that farm households in Ghana that were given formal credit increased welfare expenditure, including spending on healthcare, education, housing, sanitation and energy, whereas those that were given informal credit reduced welfare expenditures. Batini, Kim, Levine and Lotti (2010) in turn conduct a literature review on the informal economy and find mixed evidence on whether informal labor and credit markets are good or bad. Informality has been found to be negatively correlated with economic growth by lowering productivity and restricting firms and individuals' access to public services. However, within a credit-rationing framework, informal credit markets could be associated with positive growth rates by reducing the cost of credit rationing and separating the high risk firms from the low risk ones. In such settings, informal credit markets can increase efficiency and have a positive impact on the formal economy. Relevant to our study, Kislat, Menkhoff and Neuberger (2017) note that in many developing countries the advantage of informal lenders over formal lenders is in their lower collateral requirements that result from their information advantage obtained by ongoing economic and social relationships. Using household survey data from rural Thailand, they find that formal lenders rely on collateral about 40 percent more often than informal lenders, and the difference is explained by informal lenders' better information on borrowers. Shorter distance between informal lenders and borrowers improves information about the borrower. [1]

---

[1] The literature in this area is large. Dell'Anno (2015) for instance uses a model of small open overlapping-generations economy over infinite discrete time to show that multiple equilibria can exist if credit markets are imperfect and there is non-divisible entry cost in the formal economy. Doan, Gibson and Holmes (2011) in turn use survey data and propensity score matching techniques to find that microfinance has significant and positive impacts on education and healthcare for poor households in Vietnam. Moreover, the positive effects arise from access to formal credit. Field, Pande, Papp and Rigol (2013) conduct a randomized control trial in India to show that compared to a classic contract that requires immediate repayment, a contract that includes a two-month grace period increased short-run business investment, long run profit and default rates. The classic contract limits both default risk and income growth. Gine (2010) shows that rural credit market is characterized by the coexistence of formal (government and commercial banks and micro-lending institutions) and informal (moneylenders) credit. Using survey data from Thailand, the author shows that the limited ability of banks to enforce contracts is able to explain the diversity of lenders better than the transaction costs borrower face in obtaining external credit. Granda, Hamann and Tamayo (2017) develop a dynamic general equilibrium model and use survey data to show that saving constraints in the form of fixed costs to use the financial system lead households to seek informal saving instruments (cash) and result in lower aggregate saving. Hai and Heckman (2016) develop a dynamic model of schooling and work experience and use data from the National Longitudinal Survey of Youth 1997 to parameterize the model. They show that life cycle credit constraints affect human capital accumulation and inequality. The chronically poor with low initial endowments and abilities and low levels of acquired skills over the lifetime face flat lifecycle wage profiles and remain credit constrained over most of their lifetimes. The initially well-endowed persons with high levels of acquired skills has rising life cycle wage profiles. They are constrained only early on in life because they cannot immediately access their future earnings and as they age, their constraints are relaxed as they access their future earnings. Equalizing cognitive and non-cognitive ability have dramatic effects on reducing inequality in education. Chiuri and Jappelli (2001) exploit homeownership data from the Luxembourg Income Study (LIS), which consists of a collection of microeconomic data on 14 OECD countries, and find that the availability of mortgage finance as measured by outstanding mortgage loans and down payment ratios affect the age profile of home ownership, obliging young households to save and postpone home purchase until later in life. Lee and Persson (2012) argue that developing countries often have a shortage of small and medium sized enterprises, termed the "missing middle", and that this could be due to insufficient access to finance. They develop a theoretical model based on Holmstrom and Tirole (1997) to show that while family finance is prevalent among small firms and comes at giveaway prices, many borrowers are averse to it. Family finance mitigates agency problems and family investors may accept negative returns. However, borrowers may dislike family finance because they are averse of imposing risks on family or friends or were afraid of repercussions for his relationships with them. Due to these concerns, borrowers are inclined to forgo large risky investments rather than fund them through family and friends. This curtails the usefulness of family finance for entrepreneurial risk taking and growth. Nguyen and Van Den Berg (2014) conduct an empirical analysis using the Vietnam Household Living Standard Survey and find that credit from friends and relatives has a positive but not statistically significant impact on per capita expenditure. On the other hand, credit from moneylenders increases per capita expenditure of households by around 15% and reduces the poverty incidence of borrowers by around 8.5 percentage points in Vietnam. Bendig, Giesbert and Steiner (2009) use data from rural Ghana and find that poorer households are less likely to participate in the formal financial sector than better off households. Usage

There is also a literature that is relevant for our study in that it deals with labor informality in Russia. Lehmann and Zaiceva (2013) show that informal employment is a wide-spread phenomenon in the Russian labor market using panel data from the Russian Longitudinal Monitoring Survey (RLMS) between 2003-2011. The incidence of informal employment varies widely across regions in Russia, from low single digits in the high growth and diversified regions of Moscow and Sankt-Petersburg, to 23 percent in 2010 in the relatively poor Southern Region and roughly 38 percent in the North-Caucasus region. Younger workers, males, workers with primary education or less, persons with low skills, workers in construction and trade and related services have a substantially higher likelihood of being informally employed. Firm size does not capture informal employment well in an emerging economy like Russia's. Individuals who are more risk loving tend to have a higher probability to select themselves into informal employment. Slonimczyk and Gimpelson (2015) investigate the degree of persistence of informality and the extent to which informal jobs are stepping stones to a formal job by developing a dynamic multinomial logit model and using RLMS data to estimate the coefficients. Informal jobs account for about 20-25 percent of employment in Russia and persistence rates in the informal sector is almost 50 percent. The probability of transitioning to a formal job is higher if the original state is the informal sector compared to non-employment. Endogenous selection of individuals into the informal sector explain a large share of the state dependence, measured as the average difference between the probability of staying in the same labor market state and the probability of entering from other origin states, and the low transition rate into formal jobs. Once accounting for observed characteristics, state dependence of informal employee is 7.6% for males and 9.6% for females.[2]

---

of savings products, loans and insurance also depend on households' risk assessment and past exposure to shocks. Straub (2005) shows in a moral hazard framework with credit rationing that a firm's decision between the formal and informal sector is shaped by the interaction between the cost of entry into formality, and the relative efficiency of formal and informal credit mechanisms. Testing the model with firm-level data from the World Bank's 1999-2000 World Business Environment Survey, he finds that the incidence of informality is higher among smaller firms and in service and manufacturing compared to agriculture and construction. Private and local ownership increase informality. The author shows that better institutional mechanisms increase the attractiveness of formality by making market interactions more efficient. Venittelli (2016) finds that an increase in microfinance institutions (MFI) loans significantly raises the informal interest rate in the rural credit market in Andhra Pradesh. The impact of microcredit supply is positive and significant in villages with low level of competition, as measured by the landless share, but not significant in villages with high level of competition. In less competitive villages, increasing MFI loans generate detrimental consequences for the poorest, by worsening their informal credit market access conditions, while in more competitive villages, MFIs cannot subtract clients from moneylenders. Finally, Zhang, Lin and Li (2012) show that households' choice of financial intermediary is conditioned by households' social network structures and socioeconomic status using survey data from the Monitoring Economic and Social Development in the Western Regions of China (MEDOW) project. Households' social network size and network composition affect their choices by limiting the quality and quantity of information, resources, and social influence one can access through social ties. High-SES families favor formal intermediaries due not only to their richer financial knowledge, higher affordability, and greater capacity to repay loans, but also to their high demands and special types of financial needs that can hardly be satisfied by embedded resources.

[2] Clarke and Kabalina (2000) use survey data on 40 new private enterprises and over 4000 households to show that the new private sector has become an important agent in urban labor markets in Russia, accounting for between a third and a half of new hires. The new private sector employment is predominantly in small enterprises and is heavily concentrated in trade and services, with much less penetration in industry, construction and transport. Although some new private enterprises offer highly skilled professional services, in general the skill levels required by new

The paper is structured as follows. In Section 2 we present a simple theoretical model that provides predictions that we test empirically. In Section 3 we present the empirical model. In Section 4 we describe the data, while in Section 5 we discuss measures of credit market accessibility. Section 6 presents the results and policy simulations, and Section 7 concludes.

## 2. Theoretical Model

In this section, we present a simple two-period model that (a) captures the essential features of the relationship between credit market development and informality, and (b) provides testable predictions for our empirical work.

*Setup*

The link between informality and credit market accessibility can be illustrated in the standard framework that builds on Modigliani and Brumberg (1954). We abstract to a stylized two-period model for the clarity of predictions, although the extension to the infinite horizon is straightforward. Assume that a household lives for two periods - the present ($t$) and the future ($t+1$). The household faces an intertemporal budget constraint:

Period 1:    $c_t = y_t + b_t$ (1)
Period 2:    $c_{t+1} = y_{t+1} - (b_t + \kappa)(1 + r)$

The first-period budget constraint implies that the household enters period $t$ with no assets and no debt, earns income $y_t$, consumes $c_t$, and can borrow $b_t$ to cover the deficit (or save as $-b_t$). In the second period, the borrowing household pays back the loan principal $b_t$, interest $r$, and the fixed non-interest cost of borrowing $\kappa$. The saving household earns interest and does not incur cost, $\kappa = 0$. The fixed cost of borrowing may include not only the direct fees charged at closing such as origination and application fees but also indirect cost to borrowers such as the value of time spent preparing application documents and travelling to the bank. We assume that these loan application expenses are rolled into the future period payment with interest.[3]

Suppose that household income in the second period grows at a rate $g_{t+1}$:
$$y_{t+1} = (1 + g_{t+1})y_t$$
At time $t$, the household forms an expectation of future income:
$$E_t y_{t+1} = (1 + g)y_t, \quad (2)$$
where $E_t$ denotes expectation conditional on the information available in period $t$, and $g = E_t g_{t+1}$ is the expected rate of future income growth.

In the economy without borrowing constraints, we can solve the standard problem of a household that maximizes the expected utility of consumption over two periods:

---

private sector employers are not particularly high. People taking jobs in the new private sector or starting their own businesses are much more likely than others to be engaging in an activity which does not demand any of their previous skills. Many new private sector employers face rates of labor turnover even higher than those of the traditional sector. In addition, individuals who were working in traditional enterprises or organizations expressed a positive preference for working in the state sector. The prevalence of hiring through personal connections in the private sector creates a self-imposed barrier and is closely connected to the informality of employment relations and the illegality of much economic activity. It also contributes to the high earning differentials, where only those with connections have access to the better-paid jobs in the new private sector.

[3] We can also introduce other costs of borrowing as a higher interest rate, making the total payment in the second period as $b_t(1 + r + \kappa)$. Such modification produces similar predictions under slightly different assumptions.

$$U = u(c_t) + \beta E_t u(c_{t+1})$$

The utility function is assumed to be increasing, strictly concave, differentiable, and time separable, with $\beta$ as an intertemporal discount factor. The first order condition for an internal solution of the consumer's problem leads to a well-known Euler equation:

$$u'(c_t) = \beta(1+r)E_t[u'(c_{t+1})]$$

If the utility function is quadratic as in Hall (1978), $u(c_t) = -\frac{1}{2}(\check{c} - c_t)^2$, where $\check{c}$ is the bliss level of consumption, and assuming $\beta(1+r) = 1$, then the first order condition simplifies to $c_t = E_t c_{t+1}$. Using equations (1) and (2), one can solve for the desired amount of borrowing:

$$b_t^* = \frac{gy_t - \kappa(1+r)}{(2+r)}$$

The positive anticipated income growth minus future expenses, $gy_t - \kappa(1+r) > 0$, yield positive desired borrowing. The borrowing household prefers to take a larger loan (1) when the expected income growth goes up, $(b_t^*)'_g > 0$, (2) when the household earns a larger income, $(b_t^*)'_{y_t} > 0$, (3) when the interest rate falls, $(b_t^*)'_r < 0$, and (4) when other borrowing costs decrease, $(b_t^*)'_\kappa < 0$.

*Borrowing constraints*

Suppose that the household applies for a loan to one of the formal lenders and does not consider informal sources. The formal lender may set a credit limit based on the verified amount of income. It is common for lenders to verify the employment of potential individual borrowers and request documents validating the source of income. For example, Russian banks frequently ask individual loan applicants to submit the official salary account statement filled out by the employer. We assume a simple linear relation between the maximum amount of loan, $\tilde{b}_t$, and officially declared income: $\tilde{b}_t = \pi\theta y_t$, where $\theta$ is the share of income that can be verified by the lender and $\pi$ is an association parameter. The lender does not know $\theta$ and $y_t$ for each applicant but observes household's declared income $\theta y_t$. The key assumption is $\partial \tilde{b}_t/\partial \theta > 0$, or the positive association between the credit limits and the verifiable portion of income. To apply for a loan, the household must have some minimal amount of verifiable income, $\theta > 0$. If the entire household income comes from official sources, then $\theta = 1$.

With the above borrowing constraint, one can write:

$$b_t^* \leq \tilde{b}$$
$$\frac{gy_t - \kappa(1+r)}{(2+r)} \leq \pi\theta y_t \quad (3)$$

For the household facing the borrowing constraint, $\theta$ becomes an important consideration. To approach the desired amount of borrowing, individuals have incentives to increase $\theta$ by taking a job in the formal sector where income can be verified or by formalizing the labor contract with the current employer. Equation (3) suggests the following predictions under the borrowing constraint:

$$\frac{\partial \theta}{\partial \kappa} = -\frac{(1+r)}{(2+r)\pi y_t} < 0 \quad (3a) \qquad\qquad \frac{\partial \theta}{\partial r} < 0 \quad (3b)$$

The key testable implication of this model is that the credit market development associated with the reduction in borrowing costs $r$ and $\kappa$ increases the share of verifiable income and hence the share of formal employment. The prediction (3a) also implies that $\theta''_{\kappa r} < 0$ and $\theta''_{\kappa y_t} > 0$. In other words, the effect of improved credit accessibility (by lowering $\kappa$) on $\theta$ is predicted to be larger when borrowers are more credit-constrained, i.e., face a higher interest rate and earn lower income.

## 3. The Dynamics of Informality and Local Credit Market

*Timeline of the model*

Our theoretical model shows that formal sector workers benefit from the development of the credit market, and that access to credit may create additional incentives for informal sector workers to switch to a formal sector job. In Figure 1, we depict a hypothetical timeline of the job choice decision linked to the loan decision. Since most of interviews in our survey take place at the end of the year, we take the time of observation *t* to be the end of a given year. The timeline captures an individual who at time *t* does not plan to obtain a loan in the next 12 months ($P_t = 0$). The "no loan intention" condition ensures that the person's employment status at time *t* is independent of his future loan decisions.

At time *t,* an individual *i* is observed to be in one of three possible employment states *j*, $Y_{ijt}$: *F*=formal job, *I*=informal job, and *O*=no job. The individual enters the next period with a set of constant background characteristics $\tilde{X}_i$ (e.g., gender, ethnicity, and parents' education) and time-varying individual characteristics $X_{it}$ (e.g., schooling, marital status, household income, and household structure).

At time *t* or sometime after it, the individual observes characteristics of the local credit market such as proximity to bank services, $C_{i,t+u_1}$, $0 \leq u_1 < 1$, and decides to act on this information. For example, a newly opened bank office nearby may motivate the individual to apply for a loan ($P_{i,t+u_2} = 1$). The potential loan applicant knows (or learns) that formal lenders require a proof of official income.[4] An individual who does not satisfy the loan requirement (e.g., by working unofficially) may decide to switch to the formal job, $Y_{i,j=I,t} \rightarrow Y_{i,j=F,t+u_3}$, if such opportunity presents itself. We define the switch of jobs broadly, encompassing any change in the formal-informal status. It may for instance include cases when an informal employee remains at the same place of work but formalize the relationship with the current employer by signing a labor contract or by asking the employer to pay wages officially. With the documented source of labor income, the likelihood of obtaining a loan increases, and $L_{i,t+u_4} = 1$ if the loan application is successful.

At the end of the following year at time *t+1*, an econometrician observes the (new) employment status $Y_{ij,t+1}$, updated individual characteristics $X_{i,t+1}$, latest credit market conditions $C_{i,t+1}$, the respondent's intention to obtain a loan in the following 12 months $P_{i,t+1}$, and whether or not the individual has taken a loan in the past year $L_{i,(t,t+1)}$. The obvious limitation of the annual household survey is that no more than one transition in employment status per year per person may be observed. A possible chain of multiple transitions within an annual time interval (e.g., $F \rightarrow I \rightarrow O$) is recorded in the data as a single transition (e.g., $F \rightarrow O$). We denote any transition as $Y_{ijt} \rightarrow Y_{ij,t+1}$ and model it via the dynamic multinomial logit process described below.

---

[4] In Russia, different lenders have different requirements. The standard loan application package includes the borrower application, the original income statement for the last 3 or 6 months certified by the employer, and a copy of the labor book with complete records of person's official employment history. Additional documents per bank request may include employment contract as well as documents confirming other regular income such as social security benefits, property income, etc. After receiving documents from the borrower, the loan approval can take from a few hours to a few weeks depending on the type and amount of the loan.

*Dynamic multinomial logit model of employment*

Let us for the moment ignore credit-related variables and focus on the transition of individual *i* from employment status at time *t* to employment status at time *t*+1. Such transition may be modelled within a dynamic multinomial logit equation:

$$Y^*_{ij,t+1} = Y_{ijt}\gamma_j + \tilde{X}_i\beta_{1j} + X_{it}\beta_{2j} + \mu_{ij} + \varepsilon_{ij,t+1}, \qquad (4)$$

where $Y^*_{ij,t+1}$ is the latent propensity of individual *i* to be in employment status *j* at time *t*+1. The term $\mu_{ij}$ is the random effect that captures time-invariant individual-specific unobserved heterogeneity, and it is assumed to be correlated across employment states. $\varepsilon_{ij,t+1}$ is an i.i.d. error term that follows a Type 1 extreme value distribution. The process is observed at times $t = 1, \dots, T$. Individual characteristics $X_{it}$ are assumed to be strictly exogenous, that is, uncorrelated with $\varepsilon_{ij,t+1}$. To mitigate potential endogeneity concerns, the values of explanatory variables are taken at time *t*, before the transition occurs, rather than at *t*+1. The main parameter of interest is $\gamma_j$ which shows both the state dependence and the mobility of individuals across employment states between *t* and *t*+1.

Two main issues arise in the estimation of this class of dynamic models with correlated random effects. The first one is the assumption that $\mu_{ij}$ is uncorrelated with the explanatory variables. To allow for such correlation, we follow the literature by adding the Mundlak-Chamberlain device or the longitudinal average of time-varying explanatory variables, $X_{it}$ (Murtazashvili and Wooldridge, 2016; Papke and Wooldridge, 2008).

The second issue is the initial conditions problem that stems from the correlation between $\mu_{ij}$ and the initial observation $Y_{ij1}$. Endogenous initial conditions require the specification of conditional distribution for $Y_{ij1}$. In the literature, there are two common approaches to specifying this distribution. Heckman (1981) proposes to estimate equation (4) jointly with the process for the initial value of the dependent variable, as in equation (5).

$$Y^*_{ij1} = \tilde{X}_i\theta_{1j} + X_{i1}\theta_{2j} + R_{i1}\theta_{3j} + \varsigma_{ij} + \upsilon_{ij1}, \qquad (5)$$

where $\varsigma_{ij}$ is assumed to be normally distributed (with mean zero and variance of $\sigma^2_\varsigma$) and correlated with $\mu_{ij}$ but not with $\varepsilon_{ij,t+1}$. The model identification requires exclusion restrictions $R_{i1}$ or instruments for the initial employment status. These variables explain the employment status in the first observation year but not in subsequent years, i.e., they must be uncorrelated with $\varepsilon_{ij,t+1}$.

The alternative solution to the initial conditions problem is offered by Wooldridge (2005). It specifies the conditional distribution of $\mu_{ij}$ via an auxiliary model that includes the initial dependent variable $Y_{ij1}$ and a complete history of lagged explanatory variables, $X^+_i = (X_{i2}, \dots, X_{iT})$. Because our panel is unbalanced, we cannot implement this approach in its original form. Instead, we use the modified version of Wooldridge (2005) proposed by Rabe-Hesketh and Scrondal (2013), thereafter WRS. The WRS model specifies the following conditional density of the unobserved heterogeneity $\mu_{ij}$:

$$\mu_{ij} = Y_{ij1}\pi_{1j} + X_{i1}\pi_{2j} + \bar{X}_i^+ \pi_{3j} + \eta_{ij} = G_i\pi_j + \eta_{ij}, \qquad (6)$$

where $\bar{X}_i^+ = \frac{1}{T-1}\sum_{t=2}^T X_{it}$, $\eta_{ij} \sim N(0, \sigma_{\eta j}^2)$, and $Cov(\eta_{ij}, \varepsilon_{ij,t+1}) = 0$. The vector $G_i$ consists of the initial dependent variable, initial explanatory variables, and within-means of explanatory variables in subsequent periods. $\bar{X}_i^+$ is analogues to the Mundlak-Chamberlain device but without initial values. In the WRS estimator, initial conditions are not modelled separately, and thus instruments are not necessary.

Substitution of equation (6) into (4) leads to the standard random-effects multinomial logit model with a lagged dependent variable. We can re-write this model in a more conventional log-odds form by choosing the informal sector as the base category:

$$ln\frac{Pr(Y_{ij,t+1} = m, m \in \{F, O\})}{Pr(Y_{ij,t+1} = I)} = Y_{ijt}\gamma_j + \tilde{X}_i\beta_{1j} + X_{it}\beta_{2j} + G_i\pi_j + \eta_{ij} \qquad (7)$$

The WRS approach has several advantages. First, the within-means term $\bar{X}_i^+$ allows for the correlations between explanatory variables and unobserved heterogeneity $\mu_{ij}$. Second, compared to the Heckman approach, the WRS solution is less computationally intensive and does not require exclusion restrictions. Third, compared to the original Wooldridge approach, the WRS solution can be applied to unbalanced panels. For these reasons, we choose equation (7) as the base model for our analysis, although the Heckman solution to the initial conditions problem is also employed in the robustness analysis.

*Credit market and employment transitions*

In our empirical analysis, we test several hypotheses regarding the job informality and credit market development.

*Hypothesis 1. The probability of informal workers to switch to the formal job is expected to be higher for borrowers than for non-borrowers.* One way to check for the differences in the transition probabilities between borrowers and non-borrowers is via the Markov matrix of unconditional transition probabilities. We also compare the predicted transition probabilities between borrowers and non-borrowers after controlling for demographics and other explanatory variables. In addition, we employ the event study analysis to answer the following questions. Does the probability of transition to a formal job spike in the year of obtaining a loan? How does the job switching probability for borrowers change before and after the year of taking a loan?

*Hypothesis 2. Relaxing credit constraints is likely to increase the "informal → formal" transition probability.* We test this hypothesis within the dynamic framework described above by adding a two-way interaction term between the lagged employment status $Y_{ijt}$ and local credit accessibility $C_{it}$. We can think of $C_{it}$ as a reverse proxy for borrowing cost $\kappa$ in the theoretical model. A higher value of $C_{it}$ implies a closer proximity to banking services and more banks in the community, which presumably would reduce the loan-related expenses.

$$Y_{ij,t+1}^* = Y_{ijt}\gamma_{1j} + C_{it}\gamma_{2j} + (Y_{ijt} \cdot C_{it})\gamma_{3j} + \tilde{X}_i\beta_{1j} + X_{it}\beta_{2j} + G_i\pi_j + \eta_{ij} + \varepsilon_{ij,t+1} \qquad (8)^5$$

To make the interpretation of results easier, we choose the base category to be "informal jobs" in the dependent variable but "formal jobs" in the lagged dependent variable. This means that positive values of $\gamma_{3,j=I}$ or $\gamma_{3,j=N}$ in the first outcome equation for $Y_{i,j=F,t+1}$ imply an increase in the likelihood of switching to the formal sector compared to staying in the informal sector when credit accessibility improves. Thus, a positive value of $\gamma_{3,j=I}$ in the equation for formal jobs would support Hypothesis 2. The one-year lagged measure of bank availability is assumed to be exogenous conditional on the time-constant unobserved effect and time-varying observed factors that may influence the opening of new bank offices in the community. These observed factors include the lagged employment composition, demographic structure, and the market size measured by total population and household consumption. Since banks fulfill many functions (e.g., utility payments, direct deposits of pensions and salaries, and transactions with legal entities), we assume that the availability of banking services in a given year is not influenced by the future unobserved shocks to an individual's job choice, that is, $C_{it}$ is uncorrelated with $\varepsilon_{ij,t+1}$ once the proper covariates are included.

---
[5] For convenience, we keep the same notation of parameters and errors as in equation (7), although equation (8) includes additional terms.

*Hypothesis 3. The development of credit market institutions has a higher effect on reducing informality in credit constrained communities.* This hypothesis builds on the theoretical model that predicts a higher responsiveness of the formal sector income to the decrease in loan application cost for the borrowers who are more credit-constrained. We test this hypothesis by comparing the average marginal effect of $C_{it}$ on the size of the informal sector across different types of communities based on the equation (8) estimates. The communities are characterized in terms of their level of economic development and the availability of banking services. We hypothesize that the development of credit market institutions is likely to provide a greater return in the areas where previously the income growth was lower and where credit institutions were less developed.

*Hypothesis 4. The probability of obtaining a loan is expected to be higher for formal workers than for either informal workers or non-employed individuals, ceteris paribus.* This is a key assumption of our theoretical model. Even though this assumption may be self-evident from the institutional knowledge of the lending process in Russia and elsewhere, we prefer to verify it in our data. In testing this hypothesis, we use a simpler static model for the probability of taking a loan between time *t* and *t*+1.

$$L^*_{i,(t,t+1)} = Y_{ijt}\varphi_{j1} + C_{it}\varphi_2 + \tilde{X}_i\varphi_3 + X_{it}\varphi_4 + \lambda_i + v_{i,(t,t+1)},$$
$$v_{i,(t,t+1)} \sim logistic(0,1), \lambda_i \sim N(0,\sigma_\lambda^2), Cov(\lambda_i, v_{i,(t,t+1)}) = 0 \qquad (9)$$

The static model without the lagged dependent variable fits better the Russian context where the sharing of individual credit histories across financial institutions is not very common. The main parameter of interest, $\varphi_{j1}$, is expected to be negative for categories of "informal jobs" and "non-employment", with "formal jobs" being the omitted category. Since some individuals may have changed their job before time *t* in planning for future loans, we estimate equation (9) only for individuals who, at time *t*, have no intention to apply for a loan in the next 12 months.

## 4. Data

We draw on individual-, household-, and community-level data in the Russia Longitudinal Monitoring Survey – Higher School of Economics (RLMS-HSE).[6] The survey sample is a probability sample of the Russian population, and it is based on a stratified multi-stage sampling procedure. While the panel started in 1994, we use annual data for 2006-2016 when credit market questions were included in the survey. The survey is well suited for our analysis, as it contains longitudinal information about the informality status, job transitions, and credit market participation. The survey also collects community characteristics on infrastructure and markets, including the credit market, for 160 RLMS communities. These communities are located across 32 regions (which are equivalent to states in the U.S.) and all seven federal districts of the Russian Federation.[7] We restrict our analysis to prime-age individuals who are between 20 and 59 years old. In what follows, we give a detailed description of the employment variable and a brief introduction to the other variables that we use in our empirical analysis, with more details on the latter variables being provided in Appendix 1. We describe the credit-related variables in Section 5.

---

[6] The survey is done by the National Research University Higher School of Economics together with the Carolina Population Center at the University of North Carolina at Chapel Hill and the Institute of Sociology at the Russian Academy of Sciences.

*Employment status, $Y_{ijt}$*

The employment status has three broad categories: formal workers, informal workers, and the non-employed. *Formal workers* are defined as individuals who are working at "an enterprise or organization where more than one person works" and who are officially registered at their primary job. Official registration is determined based on the survey question "Are you on a work roster, written work agreement, or work contract?" *Informal workers* are comprised of (a) unregistered employees, (b) individuals who are not working at an enterprise with more than one person (they could be self-employed or hired by a private person)[8] and (c) individuals who are engaged in individual economic activity (IEA).[9] Combining these types of informal workers into one group is justifiable because their income is not easily verifiable, and their employment record is often undocumented for the loan purposes. In some specifications, we analyze the differences in both borrowing and job switching outcomes across the subgroups of informal workers. *The non-employed group* consists of individuals who do not have a primary job and are not engaged in any IEA activity in a given reference week. We do not split the non-employed into the traditional groups of unemployed and out of labor force because such distinction is of less importance for our purposes and having fewer groups also reduces the computation time.[10]

The RLMS dataset for these three employment categories goes back to 1998 and has a complete uninterrupted series over the estimation period (2006-2016). Table 1 shows the composition of employment status for every other year. There were no major structural changes over time. The share of formal sector workers fluctuated between 61 and 65 percent. The share of informal workers averaged 17 percent of the total population in the age group 20-59, or 21 percent of the employed population. The upward trend in the informal share after 2008 was largely due to the rising proportion of unregistered employees and workers hired by a private person. The share of the non-employed was about 21 percent over the sample period.

---

[7] The regions and districts are defined according to the official administrative division as of January 1, 2010. At that time, the Russian Federation was comprised of 83 regions, including the two federal cities of Moscow and St. Petersburg.

[8] The distinction between the self-employed and those working for a private person can be made in the 2006-2014 surveys, but not later.

[9] IEA workers include individuals who answered that they are currently not working but eventually admitted to being engaged in (and paid for) an individual economic activity, such as sewing, taxi driving, babysitting, selling products on the streets, etc.

[10] The unemployment rate in Russia is relatively low. It was about 5 to 7 percent in 2006-2016. If we restrict the third employment category to unemployed job seekers, the conclusions of the paper do not change.

Starting in 2008, the RLMS added a very interesting question: "What percent of earnings was received officially, that is, taxes were paid?" While there could be some doubts in the truthfulness of answers, we use this variable as an alternative definition of formal and informal workers. After dropping respondents who did not answer this question such as the IEA workers and those who worked last month but were not paid, we split the rest of the sample into three groups of earners – those with (1) only official pay, (2) partly unofficial pay, and (3) entirely unofficial pay. Of those who answered this question, 82 percent reported receiving their earnings entirely through the official channel, 11 percent admitted to getting some portion of earnings unofficially, and 7 percent acknowledged that they received all their income unofficially; see Table 1 for the time-series in these categories.

*Explanatory variables in the main specification, $\tilde{X}_i$ and $X_{it}$*

We use three different types of explanatory variables. First, there are time-invariant individual characteristics, $\tilde{X}_i$, such as gender (=1 if female), Russian ethnicity, parents' education, urban residence, community population size, and the federal district where the respondent lives.

Second, there are time-varying variables, $X_{it}$, including years of schooling, marital status, household size, number of children under the age of 14 currently living in household, and the log of real household expenditures on non-durables. The latter variable serves as a substitute for household income, which is likely to be underreported in the informal sector because of tax evasion (see Gorodnichenko *et al*., 2009). As we explained in Section 3, to mitigate the potential endogeneity of time-varying variables, we take them at time *t* instead of *t*+1 and introduce two transformations in the form of initial conditions, $X_{i1}$, and time-averaged variables, $\bar{X}_i^+$, known in the literature as a Mundlak-Chamberlain device.

Third, there are exogenous time-varying variables which belong to $X_{it}$, but they are not being transformed. These are age, age squared, calendar year effects, and the length of time between interviews. Because the transition probabilities may be sensitive to the interval between interviews, we control for the length of this interval in all dynamic estimations.[11]

*Exclusion restrictions for initial conditions, $R_{i1}$*

Heckman's (1981) solution to the initial condition problem requires exclusion restrictions (instruments) for the first-period employment status. We utilize as an instrument regional earnings at age 17 when most people in Russia finish high school. Age 17 is the most likely time of making the college/work decision in which the sectoral choice of future employment (formal vs. informal) is an important consideration. For many people, it is also the time of entry to the labor market. The data on regional earnings at age 17 goes back to year 1965 when the standard consumer price indices were not as reliable as now. To eliminate the effect of inflation, we divide regional earnings by the country mean in each year and take the log of the ratio.

---

[11] The median length between the annual interviews is 365 days, and 92 percent of all interviews are conducted within 11 to 13 months apart from each other. However, there are some extreme cases when the interval between interviews reaches as low as 7 months and as high as 17 months.

We also include a binary indicator for whether any state-owned enterprises (SOEs) were closed in the community during the 12 months preceding the first-period survey interview. We treat this variable as an exogenous shock that may influence the employment rate in the RLMS community in the first period. Finally, we control for the regional size of the government sector and informal sector at the start of the stochastic process. Both variables are measured as a percent share of regional employment.

*Estimation sample and summary statistics*

We have sequentially applied the following sample selection rules to the individuals surveyed in 2006-2016: (1) age 20-59, (2) non-missing employment status at time *t*, (3) non-missing employment status at time *t*+1, (4) non-missing control variables, and (5) the minimum of two observations per person. The last selection rule is needed to distinguish between $X_{i1}$ and $\bar{X}_i^+$. The number of observation excluded at each step is described in Table 2. After applying all the above criteria, the main estimation sample has 76,452 observations for 15,147 adults. About 53 percent of individuals in the estimation sample begin their sequence in year 2006. There was a large entry wave in 2010 when the sample was replenished (15 percent). In other years, the entry into the estimation sample is largely attributed to reaching age 20, marriage, and rejoining the panel. Due to the varying start of sequences, we control for the year of entry in every estimated equation. The panel drop-out rate is about 12 percent per year. We check the robustness of results by modelling a sample exit process in one of the specifications.

The summary statistics by employment status are presented in Table 3. Compared to formal sector workers, informal sector workers tend to be younger, male, ethnically non-Russian, less educated, raised by less educated parents, unmarried, living in larger households with more children, and located in rural and less-populated settlements. These characteristics are expected and consistent with previous studies on informality in Russia (e.g., Slonimczyk and Gimpelson, 2015; Lehmann and Zaiceva, 2015). It is worth noting that the difference in household spending on non-durable goods and services between formal and informal sector workers is less than 4 percent.[12] From the second panel of Table 3, we observe that 17-year old individuals residing in regions with below-than-average pay are more likely to work informally or to not work at all at the start of the observed employment sequence. At time *t* =1, informal sector workers tend to come from regions with a smaller share of government jobs and a larger share of informal employment. Yet, the mean differences across employment groups are not evident in terms of the incidence of living in communities with recently closed SOEs.

## 5. Credit Market Participation and Local Credit Accessibility

This section discusses credit-related variables and constructs an index of local credit market accessibility.

*Taking a loan, $L_{i(t,t+1)}$*

We measure an individual's participation in the credit market by using the survey question "Did any of household members take a loan in the last 12 months?" This question is asked annually starting in 2006. The formulation of the question in Russian ("брал в кредит") implies borrowing in the formal credit market (i.e., from financial institutions) as opposed to "брал в долг у частных лиц", which is translated as borrowing from private individuals. Almost 78 percent of all household loans reported in the RLMS over the 2006-2016 period are consumer loans for purchases of goods and services[13], 15 percent are car loans, only 6 percent are mortgages, and 1.6 percent are student loans.

---

[12] Formal workers tend to report higher income. The formal-informal unconditional gap is about 10 percent for individual average monthly labor earning and 14 percent for household disposable income.

Household borrowing from financial institutions is strongly procyclical and as can be seen in Figure 2, Russia is no exception. The share of household borrowers was 27-31 percent when the economy was growing in 2006-2008. It plunged to 17 percent during the 2009 Global Recession and then it recovered to 28 percent level by 2012. During the most recent recession (2015-2016) associated with the rise in oil prices and sanctions imposed on Russia, the share of loan-taking households fell again to a very low level of 15 percent.

From the survey question it is unclear which household member took a loan. When estimating equation (9), we select one adult member in each household to be the prime applicant for a loan using three alternative approaches. First, a random adult member is assumed to make a borrowing decision. Second, we choose the oldest male in the age group 20-59, and if the household does not have males in this age group, then we choose the oldest female. Finally, we also estimate the loan equation for the highest paid household member. This approach is the weakest of the three, as it imposes the endogenous sample selection based on employment status. In this specification, informal workers and non-working individuals have a smaller likelihood of being selected into the sample because of their lower income and larger income underreporting.

*Intention to obtain a loan, $P_{it}$*

Every year, respondents are asked about their plans to obtain a loan in the next 12 months. On average, 6.2 percent of formal sector workers, 4.2 percent of informal sector workers, and 1.3 percent of non-employed individuals intend to obtain a loan in the next 12 months. The low intention numbers do not match the actual rates of loan incidence shown in Figure 2. In other words, the planning horizon for loans appears to be very short. This fact helps in the identification, as it implies that most individuals do not plan for loans far in advance and they are hence unlikely to select their job type a year earlier in the anticipation of a loan application in the following year.

*Measures of credit market accessibility, $C_{it}$*

The unique feature of the RLMS is its community module that collects information on the proximity of bank services.[14] From this module, and with the help of several cartographic sites, we construct three measures of credit market accessibility. The first measure is a categorical indicator for the presence of banks or bank offices in the community. It takes the value of 1 if the community does not have any bank offices (Type-1), 2 if the community has only a Sberbank office (Type-2), and 3 if other banks operate in the community (Type-3). In our sample, all communities that have offices of other banks also have a Sberbank office. The second measure is the road distance to the nearest Sberbank office; it is positive for the Type-1 communities with no bank services and zero for other communities. The third measure is the road distance to the nearest bank other than Sberbank; it is zero for the Type-3 communities in which other banks operate. Sberbank is distinctively mentioned because it plays a dominant part in Russia's banking industry.[15] Because the distance data is skewed, we apply the log transformation as ln(1+distance). The specific details regarding the construction of distance measures appear in Appendix 1.

---

[13] The RLMS survey does not have information on the size of loans. According to the Russian National Bureau of Credit Histories, the average consumer loan in April 2016 was $2,300, which is about 4 months of annual earnings. https://www.nbki.ru/company/news/?id=20354&sphrase_id=95618

[14] In a different context, the distance to lender has been previously shown to be an important determinant of lending decisions by the U.S. banks (e.g., Knyazeva and Knyazeva, 2012) as well as borrowing decisions by small firms (Degryse and Ongena, 2005).

Out of the 160 RLMS communities in the 2016 survey, about 52 percent did not have any bank office, 13 percent had only a Sberbank office, and the remaining 35 percent had offices of other banks in addition to Sberbank. Figure 3A depicts the time-series of two distance measures conditional on the distance being non-zero. The average distance from communities without any bank to the nearest Sberbank office remains at about 20-21 km over the sample period. At the same time, the average distance to the nearest other bank has been trending downward from about 36 km in 2005 to 25 km in 2014. This decreasing trend is likely to reflect the increased competition in the banking sector. The scatter plot in Figure 3B indicates a strong, negative relation between the road distance to the nearest bank and the household borrowing in the formal credit market.

One issue with using these community-level measures of bank accessibility is that they do not provide any variation for medium or large cities where more than two banks operate. We supplement the community measures with regional statistics on the number of bank offices in 30 RLMS regions and two largest cities, Moscow and St. Petersburg.[16] The total number of bank offices in Russia has been steadily rising from 19,475 in 2006 to 35,850 in 2013. But then it fell to 31,576 in 2015 after the Central Bank purged undercapitalized bank institutions. The restricted access of several leading Russian banks to Western financial markets due to imposed financial sanctions may have also contributed to bank office closing in the later period. In our sample, the average number of bank offices per 1,000 persons varies from 0.08 to 0.35, with the mean of 0.2 (or two bank offices for each 10,000 persons).

Finally, we aggregate different measures of credit market accessibility into a summary index. First, we calculate a simple average of standardized *z*-scores of the following four variables: an ordered indicator for the bank presence in the community, two distance measures, and the number of bank offices per 1,000 population. Each z-score is rescaled such that a higher value means a better access to the formal credit market (e.g., shorter spatial distance to bank services or more bank offices). Then, for the convenience of interpretation, we standardize the credit market accessibility index with a mean of zero and a standard deviation of one. An alternative approach to aggregating multiple variables into a summary index is to implement principal component analysis (PCA).[17] As it turns out, the two indices – based on z-scores and PCA – are practically identical, with a simple correlation of 0.98. For no specific reason, we choose the former index in estimations.

---

[15] Sberbank (translated as Savings Bank) is a state-owned bank whose history goes back to 1841. It is the largest bank in Russia and Eastern Europe and third largest in Europe. As of 2015, it accounts for 29 percent of aggregate banking assets of Russia, includes about 100 subsidiary banks and branches, and operates over 16.5 thousand offices (http://www.sberbank.com/about).

[16] This data is published annually by the Central Bank of the Russian Federation. The number of bank offices includes bank headquarters, subsidiary credit organizations, branches, supplementary offices, and operational offices, but excludes cash offices, cash desks, and mobile cash units.

The summary statistics provided at the bottom of Table 3 illustrate that, compared to officially registered employees, individuals who are employed informally tend to live in credit-constrained communities with significantly smaller bank presence.

## 6. Results

*The probability of transition to the formal sector*

A simple way to check for a possible association between job switching and credit market participation between t and t+1 is to compare the transition matrices of employment status for borrowers and non-borrowers, as reported in Table 4. We find that compared to non-borrowers, borrowers have (i) a higher likelihood of switching from the informal sector to the formal sector (0.287 vs 0.216); (ii) a considerably higher probability of finding a formal job in t+1 if they did not have a job in t (0.199 vs 0.131); and (iii) a lower probability of losing a job if they were employed a year ago (0.036 vs 0.054 for formal workers and 0.121 vs 0.161 for informal workers). If we split informal workers into subgroups, we find that in the year when they obtain a loan, the probability of switching to the formal sector is highest for unregistered employees (0.361), followed by those hired by a private person (0.279), and IEA workers (0.288). Self-employed workers in loan-taking households tend to stay in their sector and have the lowest probability of becoming registered employees (0.186). All categories of informal workers are more likely to move to formal jobs when they participate in the credit market than if they do not participate. In other words, the first descriptive evidence seems to be indicative of a potential link between the credit market participation and leaving informality for formal jobs.

The other way to provide descriptive support of the main hypothesis is to use the event-study approach. Figure 4 shows predicted transition probability from the informal sector to formal using event-study approach. The probability model is linear and controls for each year before and after the event (the categorical timeline), a quartic polynomial of current age, and calendar year fixed effects. Event time is defined as year when the household obtains the first loan observed in the data. The plotted dots are the coefficients on the timeline variable plus the average predicted share evaluated at timeline=0 (which is the omitted category). Panel A depicts the predicted rate of entry of informal workers to the formal sector as a share of formal workers in t. In the year of obtaining the loan, there is a global spike that shows the increased share of formal sector workers. Other local spikes do not have empirical explanation. Panel B shows the predicted switching probability for informal sector workers at period t-1 to move to the formal sector in period t. There exists a global maximum in the predicted probability in year t when the person takes a loan. There is also an increase in the switching probability between t-2 and t-1. This could mean that agents are preparing for loan application in advance. However, the previous discussion of the intention to apply for a loan reveal that planning horizon is relatively short.

---

[17] The first component accounts for 70 percent of the total variance, with the following factor loadings: an indicator for bank presence (0.586), distance to the nearest Sberbank office (-0.556), distance to the nearest office of other banks (-0.562), and the number of bank offices per population (0.178).

*Results of the main model*

Results of the main model are reported in Table 5. Table presents the relative risk ratios from the dynamic multinomial logit model of employment with unobserved heterogeneity. The dependent variable is employment status in t+1, with the informal status chosen as the base outcome and the formal sector job chosen as the omitted lagged dependent category. Column 1 reports the mean and standard deviation of variables in the estimation sample. Columns 2 and 3 present the full estimates of equation (8) with correlated random effects and using the WRS solution to the endogeneity of initial conditions.

Previous labor status plays an important role in the type of the current job. Predictably, the risk of finding a job in the formal sector relative to the probability of finding a job in the informal sector is small for both informal sector workers and non-employed (the odds are 0.155 and 0.169, respectively[18]). Results of the main interest are the two-way interactions between the credit market accessibility index and labor market status in previous period. For a formal sector worker, a one-standard-deviation increase in the credit market accessibility index increases the odds of staying in the formal sector relative to switching to the informal sector[19] by 1.253 times (or 25 percent). At the same time, for a formal sector worker, a one-standard-deviation improvement in $C_{it}$ increases the relative risk of becoming non-employed relative to finding an informal sector job[20] by 1.154 times. It could be because the informal sector shrinks. Theoretically, it is also possible that more people are becoming non-employed. For example, if the improved credit accessibility provides incentives for firms to formalize, then larger labor cost from higher taxes and more regulation may lead to layoffs. To see which effect dominates, we calculate the average marginal effect (AME) of $C_{it}$ on the probability of switching between employment states. These results are reported at the bottom of Table 5. We see that an increase in $C_{it}$ by one standard deviation reduces the probability of losing a formal sector job next period by a slight margin (-0.2 percentage points or ppt). Thus, we do not find an adverse effect of the credit market development on employment. But the effect of $C_{it}$ on the likelihood of staying in the formal sector is substantial (1.7 ppt improvement). Similarly, there is a large negative effect on the probability of moving to the informal sector (-1.5 ppt). The relative risk ratios above one for the interaction terms in the first outcome equation (1.215 and 1.228 for informal sector workers and non-employed individuals, respectively) imply that the effect of local credit accessibility on corresponding odds is even stronger for these groups than for formal sector workers. And if we look at the AME results, we can see that improved credit market accessibility (by one standard deviation) increases the chances of informal sector workers to formalize by 5.4 ppt. Their likelihood of staying in the informal sector goes down by 3.9 ppt and the probability of losing a job drops by 1.4 ppt per unit increase in $C_{it}$. The average marginal effects of in $C_{it}$ for non-employed individuals are smaller in magnitude, but they are still substantial. The transition

---

[18] These effects that can be interpreted for the communities with mean zero credit market accessibility index.

[19] The odds of staying in the formal sector relative to switching to the informal sector is estimated by the following formula: $\frac{Pr(Y_{i,t+1}=F|Y_{i,t}=F)}{Pr(Y_{i,t+1}=I|Y_{i,t}=F)}$

[20] The relative risk of becoming non-employed relative to finding an informal sector job is estimated by the following formula: $\frac{Pr(Y_{i,t+1}=N|Y_{i,t}=F)}{Pr(Y_{i,t+1}=I|Y_{i,t}=F)}$

probability from non-employed status to the formal goes up by 3.9 ppt and probability from non-employed status to the informal reduces by 3.6 ppt.

We also calculate the average marginal effect of $C_{it}$ on the size of sectors. The one-unit improvement in credit market accessibility index increases the population share of the formal sector by 2.3 ppt and reduces the informal sector by 2.1 ppt. The effect of $C_{it}$ on non-employment is negative but small in magnitude (-0.2 ppt). In addition to average marginal effects (which are calculated across all individuals in the sample), we also calculate marginal effects at the mean values of covariates and at different values of $C_{it}$. These results are plotted in Figure 5. Figure shows the predicted probabilities of being in one of the three employment states at different values of the credit market accessibility index. Results show a rise in the size of the formal sector and a decline in the size of the informal sector, as credit market accessibility improves. The employment rate appears to be unaffected.

Table 5 presented results of other variables that could be worth mentioning. Older individuals are less likely to be non-employed than work informally. Females have a much higher likelihood of being non-employed and working formally than working informally. Ethnic Russian tend to work in the formal sector. More educated individuals are more likely to work in the formal sector and less likely to be non-employed. Time-varying variables appear as time-averaged variables and as deviations from the mean. For example, non-employed individuals come from low-consumption households (0.526), but an increase in consumption above the mean raises the likelihood of being non-employed (1.068), and it can be interpreted as an income effect. Individuals have an increased risk of being non-employed with more children. Larger cities tend to have a larger informal sector. These results are consistent with summary statistics by sector.

*Robustness analysis*

The robustness analysis of the dynamic employment model of employment is presented in Table 6. Table reports the results on main variables from seven alternative specifications: specification (1) excludes unobserved individual heterogeneity and assumes exogenous initial conditions; specification (2) includes unobserved heterogeneity, but excludes time-averaged variables and initial conditions; specification (3) uses the Heckman solution to the endogeneity of initial conditions[21]; specification (4)[22] includes additional controls such as real regional GDP per capita (in log), the regional unemployment rate, the 5-year moving average of the regional inflation rate, the log of distance from the community center to the regional capital, and a dummy for whether any state-owned enterprises were closed in the community in the last 12 months[23]; specification (5) excludes the city of Moscow[24]; specification (6) re-estimates equation (8) on the sample of individuals who at time t did not plan to take a loan in the next 12 months; specification (7) adds "leaving the survey in t+1" as a fourth possible individual outcome.

---

[21] The full Heckman estimates including equations for initial conditions with exclusion restrictions are shown in Appendix Table A2.1

[22] Specifications (4)-(7) use the WRS solution to the endogeneity of initial conditions and have the same set of covariates as in Table 5.

[23] The model is computationally time-consuming, and adding additional variables is costly in terms of time. We are confident in the variable choice in our main specification but also check if the results change when additional regional/community controls are added.

[24] Almost 50 percent of country's credit organizations and 10 percent of all bank offices are in Moscow. Given such a large bank concentration in one city, we check if our results stay when we exclude Moscow respondents from the sample.

Robustness analysis supports results of the main model presented in Table 5. From these results we can conclude that it is important to account for individual time-constant unobserved heterogeneity, as results change substantially. The assumption about exogeneity of initial conditions could be a strong assumption[25], the models that account for the endogeneity of initial conditions produce somewhat larger effects of credit market accessibility on the size of informal and formal sectors, but the overall story is the same. Results show that there is no change from adding additional controls or excluding the city of Moscow. Also, for the sample of individuals who at time t did not plan to take a loan in the next 12 months, the effect of credit market accessibility on the formal and informal sector proportions is only slightly larger than for the full sample by 0.1 ppt. Accounting for the survey exit does not make much difference to the main result but reduces the computation time by almost 40 times. It is interesting that informal workers and the non-employed have a lower likelihood of survey exit relative to staying in or switching to the informal sector.

Results show that Heckman and WRS solutions to the endogeneity of initial condition produce similar effects. The full Heckman estimates including equations for initial conditions with exclusion restrictions are shown in Appendix Table A2.1. This table provides also results of the model with exogenous initial conditions. Specification 1 assumes exogenous initial conditions. The comparison of this specification with the WRS estimator in Table 5 shows that the results on interaction terms related to job switching are not that different. Some other coefficients are statistically different such as age and household size, although the direction of the effect is the same between the two specifications. Results suggest that the exogeneity of initial conditions might be a strong assumption. Specification 2 provides results for the Heckman specification. The comparison of this specification with the WRS specification in Table 5 shows that the main results are very similar. The increase in credit market accessibility index by one standard deviation significantly increases the risk of staying in the formal sector compared to moving to the informal sector. Similarly, for informal workers, the likelihood of finding a formal job goes up as $C_{it}$ increases. The only result that is different between the Heckman and WRS estimator is the age effect in the first outcome equation. In the WRS model, the age effect is flat, while in the Heckman model, it is concave (increasing at first and then declining at later age). Exclusion restrictions for first-period employment status are statistically significant. Predictably, we find that individuals in regions with a larger government sector and smaller informal sector are more likely to have a formal job in the first period of their employment sequence. Recently closed SOEs in the community increase the probability of being without a job. Individuals who went to high school in the regions with above-than-average pay are more likely to work formally in the first period. But they are also likely to be non-employed than to have an informal job. This result is less intuitive and might be due to the income effect on work incentives. Generally, we prefer the WRS solution since it does not rely on having exclusion restrictions.

Other way to provide the robustness analysis is consideration of disaggregated employment statuses. The lagged employment status at time t is disaggregated by splitting informal workers at time t into the four subgroups. In all other ways, the two specifications in this table are the same as in Table 5. For the communities with mean zero credit market accessibility index, all groups of informal workers have a lower risk of moving to the formal sector relative to staying in the informal sector. This relative risk is the lowest for the self-employed (0.111), followed by IEA workers (0.168), those hired by a private person (0.184), and then unregistered employees (0.196). It is interesting that the relative risk of losing a job is also the lowest for the self-employed (0.279), which is consistent with results in Table 4. Similar to Table 5, a one-standard-deviation increase in the credit market accessibility index increases the odds for formal sector workers to remain in the same sector relative to switching to the informal sector[26] by 1.229 times (or 23 percent). The AME results show that all subgroups in the informal sector have a large positive effect of the credit market accessibility index on the probability of switching to the formal sector job in the next period. The effect varies between 2.7 ppt for self-employed and 5.5 ppt for IEA workers for one-unit increase in $C_{it}$. There is less heterogeneity in the negative effect of $C_{it}$ on the probability of remaining in the informal sector. This probability declines by 3.4-4.9 ppt for every unit increase in $C_{it}$. Finally, the likelihood of becoming non-employed is unaffected by the credit market development for all subgroups but the self-employed. This result is possible for several reasons. First, it could be just an income effect. For example, receiving a loan by a household member may reduce incentives of other members being engaged in self-employment. Second, some self-employed may lose their job if more banks mean more regulation and stronger tax enforcement in the community. The puzzle here is why only self-employed are affected, but not other groups.

---

[25] The full exogenous initial conditions model estimates are shown in Appendix Table A2.1

The other possible definition of the informal labor status is based on the type of payment. Table A2.2 provides estimates for the payment type instead of registration status. The payment type has four categories: (1) only official pay, (2) partly unofficial pay, (3) entirely unofficial pay, and (4) no job. IEA workers and individuals with no labor earnings in the last 30 days are excluded because there is no information on their payment status. At mean zero credit market accessibility index, workers with only unofficial pay have significantly lower odds of moving to jobs with only official pay compared to workers with partly unofficial pay (0.052 vs 0.491). However, these odds rise considerably when the borrowing opportunities expand. The average marginal effect of $C_{it}$ on the probability of receiving earnings officially is positive for all four groups, but it is especially large for those who previously received their earnings only unofficially, in so called envelopes (6.8 ppt). With credit market development, this group is less likely to remain in the same status by 4.2 ppt per every unit increase in the index. It could be noted that workers with partly unofficial pay can still get a loan based on the declared portion of their income. This could explain why the effect of credit market accessibility on the transition probabilities for this group is not as large as the one for workers with only unofficial pay.

*Policy simulations*

For policy simulations, we simulate separate components of credit market accessibility index. Table A2.3 provides results of the WRS model for different measures of credit market accessibility. We replace the aggregate index of credit market accessibility by the vector of index components. The disadvantage of this approach is having too many interactions. This makes the direct interpretation of relative risk ratios more difficult. Yet, the advantage of such replacement is the ability to simulate policies based on specific indicators rather than on index.

---

[26] The odds for formal sector workers to remain in the same sector relative to switching to the informal sector is estimated by the following formula: $\frac{\Pr(Y_{i,t+1}=F|Y_{i,t}=F)}{\Pr(Y_{i,t+1}=I|Y_{i,t}=F)}$

We report the summary results of four simulated policies in Table 8. Table 8 shows the predicted probabilities of being in the formal and informal sectors before and after a given policy is implemented. The predicted probability of being non-employed is $1 - \widehat{P}_F - \widehat{P}_I$, and it is not shown. Predicted probabilities are evaluated at the starting values indicated in the table and at sample means of all other covariates. Thus, these are not average marginal effects across the sample, but the marginal effects at given values.

The first policy is opening a Sberbank office in the community that does not have any bank within 10 km distance. Based on the model estimates, we find a statistically significant increase in the size of the formal sector from 63 percent to almost 65 percent as a result of this policy. The informal sector falls by 1 ppt, and the share of jobless individuals falls by 0.9 ppt.

Now, suppose that instead of Sberbank, some other bank opens its office in the community where no other bank operated previously. The effects of this policy are larger in magnitude. The estimates predict a substantial decline in the share of informal sector by 3.9 ppt. The formal sector expands by 2.9 ppt, but there is a negative spillover effect of rising non-employment by 1 ppt. We already mentioned a few reasons for why such side effect may occur. First, more regulation and higher taxes raise costs for firms and may lead to job losses. Second, credit market development may generate additional income effect that may reduce work incentives.

The third policy is opening a second bank in the community that already has a Sberbank office. We also find a substantial effect of bank competition on the size of the informal sector. For the community where more than two banks operate, we change the regional number of banks per 10,000 from 2 to 3. This policy is also predicted to have a substantial effect on employment composition– increase in the share of the formal sector by 2 pp and decrease in the share of the informal sector by 2.6 ppt.

Overall, policy simulations show a strong support for the reduction in informal employment in response to better credit market accessibility.

*The effect of credit market accessibility on the informal sector*

Table 9 shows the average marginal effect of credit market accessibility on the size of the informal sector by subgroups of individuals as well as the predicted mean size of the informal sector. The gender differences are shown to motivate future research, as we find that males are not only more likely to work in the informal sector, but they are also more responsive to changes in credit market accessibility.

Other results intend to test one of the predictions of the theoretical model - the effect of lifting credit constraints is larger for credit-constrained communities with lower levels of income. Despite that these communities may have lower job opportunities in the formal sector, their informal sector share is more responsive to the development of credit market. The treatment effect is estimated to be higher for regions with lower average earnings and higher unemployment rate and for communities with no banks and with a lower index of credit market accessibility, as the theoretical model predicts.

*Loan equation*

In the table 10, we provide results for the household's probability of taking a loan between time t and t+1 and for the specific types of loans. Mortgage and auto loans require collateral such as a purchased car or a house. Consumer loans do not require a collateral, but they do require a proof of income. Both models are estimated on the sample of individuals who at time t did not plan to take a loan in the next 12 months. An increase in the credit market accessibility index by one standard deviation increases the likelihood of obtaining a loan by 3.5 percentage points or by 1.32 times using the ratio of odds. Mortgage and auto loans are less responsive to the availability of bank services compared to consumer loans. opening a bank office in the community does not immediately qualify potential borrowers for a mortgage or a car loan. The decision on house or car purchase is more deliberate and require more efforts, preparation of documents and time consuming, and the distance to the nearest bank might be less important consideration for potential borrowers in this case.

Informal sector workers and people without a job are less likely to receive a loan relative to formal sector workers. This supports the hypothesis that the probability of obtaining a loan is expected to be higher for formal sector workers than for the other two groups. The relative risk ratio is 0.842 and 0.642 for informal sector workers and the non-employed, respectively. Informal sector workers have a much lower likelihood of obtaining a mortgage and a car loan compared to a consumer loan, which is expected.

Among other expected results it should be mentioned that the likelihood of receiving a loan is higher for females, ethnic Russians, married, living in households with more kids, and households with higher levels of consumption. Younger people have higher chances of obtaining a consumer loan, maybe because many older people were not used to taking loans in the past economic system and often borrowed from each other. Other interesting results: gender does not matter for mortgage and car loans; more educated individuals have an increased risk by 3.5% of taking a mortgage and auto loan, but a decreased risk of taking a consumer loan by 7%; household size matters for the mortgage and car loans but not for consumer loans; the opposite situation with number of kids: having more kids does not increase the likelihood of mortgages and car loan but improves the odds of obtaining consumer loans. Results show existence of the lower probability of taking a consumer loan in more populated area. It can be explained by higher competition among borrowers. But it also could be that people in more populated communities are less credit constrained and can buy goods and services directly without borrowing.

Table A2.4 provides results for loan equations with alternative definitions of the head of household: (1) a random household member, (2) oldest male or oldest female if there are no males in household, and (3) a highest earning member of household. No matter how we choose the head of household, the risk of taking a loan increases when the credit market accessibility increases, and it is significantly lower for informal sector workers and the non-employed compared to formal sector workers.

## 7. Conclusions

The study investigates a novel mechanism of reducing the labor informality through the development of the credit market. In this paper, we provide a simple two period theoretical model and an empirical estimation of the link between the credit market development and the labor mobility among three labor market states (informal sector, formal sector and non-employment). The theoretical model shows that the credit market development associated with a reduction in interest costs and non-interest costs of borrowing increases the share of formal employment, and the empirical work tests this evidence.

The main empirical method is the dynamic multinomial logit model of employment with correlated random effects. Two main issues arise in the estimation of this class of dynamic models with correlated random effects. The first one is the assumption that the error term is uncorrelated with the explanatory variables. To allow for such correlation, we follow the literature by adding the Mundlak-Chamberlain device or the longitudinal average of time-varying explanatory variables. The second issue is the initial conditions problem that stems from the correlation between the error term and the initial observation for the employment status. Endogenous initial conditions require a specification of the conditional distribution for the initial employment status. In the literature, there are two common approaches to specifying this distribution: Heckman (1981), and Wooldridge (2005) and Rabe-Hesketh and Scrondal (2013). The paper estimates and compares both solutions to the initial conditions problem.

The paper shows that the probability that informal workers would switch to formal jobs is higher for borrowers than for non-borrowers. Furthermore, a relaxation of credit constraints increases the probability of transition from an informal to a formal job. These results are robust in different specifications of the model. Policy simulations show a strong support for a reduction in the informal employment in response to better CMA in credit constrained communities.

# Tables

## Table 1. Trends in Employment Status

|  | Mean | 2006 | 2008 | 2010 | 2012 | 2014 | 2016 |
|---|---|---|---|---|---|---|---|
| Formal job | 0.620 | 0.620 | 0.645 | 0.628 | 0.609 | 0.610 | 0.614 |
| Informal worker | 0.168 | 0.165 | 0.149 | 0.151 | 0.179 | 0.180 | 0.180 |
|    Unregistered employee | 0.042 | 0.046 | 0.036 | 0.041 | 0.044 | 0.046 | 0.048 |
|    Self-employed | 0.024 | 0.019 | 0.026 | 0.023 | 0.025 | 0.026 | … |
|    Works for a private person | 0.046 | 0.034 | 0.036 | 0.042 | 0.056 | 0.057 | … |
|    IEA worker | 0.051 | 0.064 | 0.051 | 0.044 | 0.053 | 0.049 | 0.057 |
|    Unknown | 0.002 | 0.002 | 0.001 | 0.001 | 0.002 | 0.002 | … |
| No job | 0.212 | 0.215 | 0.206 | 0.221 | 0.212 | 0.210 | 0.207 |
| Number of observations | (109,957) | (8,247) | (7,879) | (12,044) | (12,362) | (9,920) | (9,730) |
| Formal pay only | 0.825 | … | 0.842 | 0.830 | 0.851 | 0.835 | 0.799 |
| Some unofficial pay | 0.107 | … | 0.111 | 0.116 | 0.093 | 0.096 | 0.101 |
| Unofficial pay only | 0.068 | … | 0.048 | 0.054 | 0.056 | 0.069 | 0.101 |
| Number of observations | (57,243) | … | (4,395) | (7,033) | (7,289) | (6,006) | (7,103) |

**Notes**: Table shows the composition of the surveyed population of age 20-59 for all available years combined starting with 2006 (column 1) and separately for every other survey year. The definition of each status is provided in Appendix 1.

## Table 2. Sample Selection Criteria

| Sample selection criteria | Number of person-year observations | |
|---|---|---|
|  | Excluded | Remaining |
| RLMS, 2006-2016, age 20-59 |  | 115,521 |
|    *Less* |  |  |
| Missing employment status in *t* | 5,564 | 109,957 |
| Missing employment status in *t*+1 | 29,496 | 80,461 |
| Missing information on explanatory variables | 193 | 80,268 |
| Only one valid observation person | 3,816 | 76,452 |

**Notes**: The main estimation sample has 76,367 person-year observations. Some estimation samples may have a smaller number of observations due to additional sample constraints.

**Table 3. Summary Statistics**

|  | *Formal workers* | *Informal workers* | *Non-employed* |
|---|---|---|---|
| Age | 39.706 | 38.208 | 40.670 |
|  | (10.555) | (10.749) | (13.402) |
| Female (binary) | 0.532 | 0.414 | 0.658 |
| Russian ethnicity (binary) | 0.884 | 0.788 | 0.803 |
| Parents' education |  |  |  |
|    Lower levels | 0.474 | 0.528 | 0.565 |
|    Upper vocational | 0.247 | 0.211 | 0.198 |
|    Higher education | 0.199 | 0.171 | 0.168 |
|    Missing | 0.079 | 0.090 | 0.068 |
| Years of schooling | 12.391 | 11.203 | 11.081 |
|  | (2.282) | (2.196) | (2.421) |
| Married (binary) | 0.622 | 0.527 | 0.523 |
| Number of HH members | 3.440 | 3.749 | 3.719 |
|  | (1.433) | (1.770) | (1.860) |
| Number of kids per HH, 0-13 | 0.607 | 0.650 | 0.584 |
|  | (0.782) | (0.853) | (0.878) |
| Real HH non-durable consumption | 44.976 | 43.256 | 37.077 |
|    (2016 thousand rubles) | (37.833) | (38.378) | (36.035) |
| Community population (thousand people) | 1241.568 | 909.508 | 1105.286 |
|  | (3035.108) | (2551.442) | (2962.707) |
| Urban (binary) | 0.765 | 0.657 | 0.633 |
| Interval between interviews (days) | 363.929 | 364.416 | 363.935[n] |
|  | (26.116) | (24.702) | (25.192) |
| *Estimation sample* | [47,468] | [12,732] | [16,252] |
| Regional earnings at age 17 relative to | -0.053 | -0.105 | -0.103 |
|    country mean, log difference | (0.225) | (0.244) | (0.268) |
| Government sector size, % of employment | 30.825 | 30.601 | 30.859[n] |
|  | (4.335) | (4.120) | (4.315) |
| Informal sector size, % of employment | 18.322 | 21.166 | 20.903 |
|  | (8.008) | (9.238) | (9.853) |
| Any SOEs closed last 12 months? (binary) | 0.202 | 0.201[n] | 0.214[n] |
| *Initial conditions sample (t=1)* | [9,106] | [2,468] | [3,573] |
| Credit market accessibility index | 0.096 | -0.187 | -0.238 |
| Bank presence in the community | (0.902) | (1.137) | (1.140) |
|    No banks | 0.139 | 0.231 | 0.254 |
|    Only Sberbank | 0.065 | 0.068[n] | 0.074 |
|    Other banks | 0.796 | 0.701 | 0.672 |
| Distance to the nearest Sberbank office, km | 2.731 | 5.463 | 5.380 |
|  | (8.710) | (12.124) | (11.397) |
| Distance to the nearest other bank office, km | 5.592 | 9.135 | 9.724 |
|  | (9.135) | (9.724) | (13.732) |
| Number of bank offices per 1000 population | 0.209 | 0.198 | 0.199 |
|  | (0.198) | (0.199) | (0.061) |
| *Estimation sample* | [47,468] | [12,732] | [16,252] |

**Notes**: This table shows the mean and standard deviations of variables used in the empirical analysis. The standard deviations of binary variables are not reported. The number of observations is in brackets. The definition of each variable is provided in Appendix 1. HH denotes household. All unconditional differences between formal and informal workers and between formal employees and the non-employed are statistically significant at the 5 percent level, except for the estimates marked with a superscript "*n*".

**Table 4. Average Transition Probabilities for Borrowers and Non-Borrowers**

| Status j in t | Probability of transition to the formal job in t+1, $P_{jF}$ | | Probability of transition out of employment in t+1, $P_{jN}$ | |
|---|---|---|---|---|
| | Borrowers [t, t+1] | Non-borrowers [t, t+1] | Borrowers [t, t+1] | Non-borrowers [t, t+1] |
| *t∈[2006-2015]* | | | | |
| Formal job | 0.900 | 0.891 | 0.036 | 0.054 |
| Informal job | 0.287 | 0.216 | 0.121 | 0.161 |
| No job | 0.199 | 0.131 | 0.655 | 0.741 |
| Number of observations | (12,124) | (35,455) | (2,588) | (13,456) |
| *t∈[2006-2013]* | | | | |
| Unregistered employee | 0.361 | 0.322 | 0.091 | 0.129 |
| Self-employed | 0.186 | 0.175 | 0.062 | 0.057 |
| Works for a private person | 0.279 | 0.253 | 0.076 | 0.107 |
| IEA worker | 0.288 | 0.151 | 0.248 | 0.293 |
| Number of observations | (10,566) | (27,503) | (2,228) | (10,585) |

**Notes**: This table shows the average annual probabilities of (i) transitioning from status *j* at time *t* to the formal sector in *t*+1, $P_{jF}$ and (ii) exiting from employment between *t* and *t*+1, $P_{jN}$. The probability of staying in or switching to informal job is $1 - P_{jF} - P_{jN}$ (not shown). The transition probabilities are calculated separately for borrowers and non-borrowers, which are defined based on the survey question asked in *t*+1: "Did any of household members take a loan in the last 12 months?" Sample is limited to prime-age individuals (age 20-59). The definition of each status is provided in Appendix 1.

## Table 5. Dynamic Multinomial Logit Model of Employment Choice

|  | Mean (SD) | Formal job, t+1 | No job, t+1 |
|---|---|---|---|
|  | 1 | 2 | 3 |
| Credit market accessibility index, $C_{it}$ | -0.022 | 1.253*** | 1.157*** |
|  | (1.010) | (0.061) | (0.063) |
| Informal job, t | 0.167 | 0.155*** | 0.695*** |
|  | (0.373) | (0.009) | (0.044) |
| No job, t | 0.213 | 0.169*** | 2.801*** |
|  | (0.409) | (0.010) | (0.177) |
| (Informal job, t) x $C_{it}$ | -0.031 | 1.215*** | 0.968 |
|  | (0.469) | (0.053) | (0.046) |
| (No job, t) x $C_{it}$ | -0.051 | 1.228*** | 1.119** |
|  | (0.534) | (0.056) | (0.052) |
| *Control variables, $\tilde{X}_i, X_{it}$* | | | |
| Age, t | 39.661 | 0.997 | 0.739*** |
|  | (11.277) | (0.016) | (0.013) |
| Age squared /100, t | 17.002 | 1.002 | 1.549*** |
|  | (9.032) | (0.020) | (0.034) |
| Female | 0.539 | 1.503*** | 2.699*** |
|  | (0.498) | (0.071) | (0.136) |
| Russian ethnicity | 0.850 | 1.462*** | 1.078 |
|  | (0.357) | (0.096) | (0.074) |
| Parents' education | | | |
|    Upper vocational | 0.231 | 1.013 | 1.142** |
|  | (0.421) | (0.061) | (0.074) |
|    Higher education | 0.188 | 0.982 | 1.279*** |
|  | (0.391) | (0.067) | (0.095) |
|    Missing | 0.079 | 0.685*** | 0.750*** |
|  | (0.269) | (0.065) | (0.077) |
| Years of schooling, t | 11.914 | 1.127*** | 0.928** |
|  | (2.378) | (0.037) | (0.035) |
| Married, t | 0.585 | 1.148* | 1.115 |
|  | (0.493) | (0.096) | (0.105) |
| Number of HH members, t | 3.551 | 1.005 | 0.968 |
|  | (1.598) | (0.028) | (0.030) |
| Number of kids per HH, 0-13, t | 0.609 | 1.014 | 1.114** |
|  | (0.816) | (0.049) | (0.058) |
| Real HH consumption, log, t | 3.507 | 0.996 | 1.065 |
|  | (0.722) | (0.038) | (0.042) |
| Community population, log | 4.213 | 0.928*** | 0.963** |
|  | (2.939) | (0.014) | (0.015) |
| Urban | 0.719 | 0.994 | 0.918 |
|  | (0.449) | (0.094) | (0.092) |
| Interval between interviews, t | 364.011 | 1.001** | 1.000 |
|  | (25.691) | (0.001) | (0.001) |
| *Time-averaged variables, $\bar{X}_i^+$ (Mundlak-Chamberlain device)* | | | |
| Years of schooling | 11.952 | 1.292*** | 1.008 |
|  | (2.353) | (0.052) | (0.041) |
| Married | 0.590 | 1.112 | 1.069 |
|  | (0.465) | (0.118) | (0.128) |
| Number of HH members | 3.542 | 0.963 | 1.272*** |
|  | (1.509) | (0.037) | (0.053) |
| Number of kids per HH, 0-13 | 0.610 | 1.052 | 0.927 |
|  | (0.754) | (0.070) | (0.067) |
| Real HH consumption, log | 3.521 | 1.074 | 0.526*** |
|  | (0.609) | (0.070) | (0.035) |

|  | Initial conditions, $Y_{i1}$ and $X_{i1}$ | | |
|---|---|---|---|
| Informal job | 0.165 | 0.067*** | 0.302*** |
|  | (0.371) | (0.005) | (0.025) |
| No job | 0.223 | 0.204*** | 2.377*** |
|  | (0.416) | (0.016) | (0.197) |
| Years of schooling | 11.760 | 0.804*** | 1.047 |
|  | (2.331) | (0.027) | (0.040) |
| Married | 0.568 | 1.064 | 1.094 |
|  | (0.495) | (0.095) | (0.105) |
| Number of HH members | 3.599 | 0.977 | 0.932** |
|  | (1.499) | (0.030) | (0.030) |
| Number of kids per HH, 0-13 | 0.623 | 0.955 | 0.889** |
|  | (0.785) | (0.050) | (0.049) |
| Real HH consumption, log | 3.411 | 0.947 | 1.097** |
|  | (0.789) | (0.040) | (0.047) |
| $Var(\hat{\eta}_i)$ |  | 2.657 | 2.342 |
|  |  | (0.134) | (0.126) |
| $Cov(\hat{\eta}_i^F, \hat{\eta}_i^N)$ |  | 1.379 |  |
|  |  | (0.108) |  |

*Post-estimation predictions: Average marginal effect of $C_{it}$*

|  | Employment status in t+1 | | |
|---|---|---|---|
|  | **Formal job** | **Informal job** | **No job** |
| Sector size in *t+1* | 0.023 | -0.021 | -0.002 |
| Transition probability |  |  |  |
| $F_t \rightarrow$ | 0.017 | -0.015 | -0.002 |
| $I_t \rightarrow$ | 0.054 | -0.039 | -0.014 |
| $O_t \rightarrow$ | 0.039 | -0.036 | -0.002 |

**Notes**: N=76,452. Table presents the relative risk ratios from the dynamic multinomial logit model of employment with unobserved heterogeneity. The dependent variable is employment status in *t+1*, with the informal status chosen as the base outcome. Column 1 reports the mean and standard deviation of variables in the estimation sample. Columns 2 and 3 present the full estimates of equation (8) with correlated random effects and using the WRS solution to the endogeneity of initial conditions. The following variables are included but not shown: year dummies, seven federal districts, fixed effects for the first year of the stochastic process, and the intercept. The omitted category of parents' education is "general secondary education or below". Robust standard errors clustered by individual id are in parentheses; *** $p<0.01$, ** $p<0.05$, * $p<0.1$. Variables without subscript *t* are time-constant. The average marginal effects are calculated across individuals in the estimation sample.

**Table 6. Robustness Analysis of the Dynamic Employment Model with Unobserved Heterogeneity**

|  | Reduced-form (1) | | Exogenous initial conditions (2) | | Heckman estimator (3) | | WRS estimator, additional controls (4) | |
|---|---|---|---|---|---|---|---|---|
|  | Formal, t+1 | No job, t+1 | Formal, t+1 | Formal, t+1 | No job, t+1 | No job, t+1 | Formal, t+1 | No job, t+1 |
| Credit market accessibility index, $C_{it}$ | 1.106*** | 1.025 | 1.182*** | 1.078 | 1.284*** | 1.137** | 1.276*** | 1.213*** |
|  | (0.038) | (0.042) | (0.051) | (0.054) | (0.062) | (0.062) | (0.066) | (0.070) |
| Informal job, t | 0.029*** | 0.309*** | 0.055*** | 0.595*** | 0.155*** | 0.694*** | 0.155*** | 0.693*** |
|  | (0.001) | (0.014) | (0.004) | (0.039) | (0.009) | (0.042) | (0.009) | (0.044) |
| No job, t | 0.081*** | 5.681*** | 0.078*** | 4.499*** | 0.178*** | 2.775*** | 0.168*** | 2.762*** |
|  | (0.003) | (0.246) | (0.004) | (0.255) | (0.010) | (0.170) | (0.010) | (0.175) |
| (Informal job, t) x $C_{it}$ | 1.202*** | 0.958 | 1.213*** | 0.950 | 1.215*** | 0.966 | 1.210*** | 0.969 |
|  | (0.048) | (0.038) | (0.057) | (0.045) | (0.053) | (0.046) | (0.053) | (0.046) |
| (No job, t) x $C_{it}$ | 1.282*** | 1.204*** | 1.271*** | 1.180*** | 1.243*** | 1.115** | 1.230*** | 1.129*** |
|  | (0.048) | (0.046) | (0.057) | (0.054) | (0.057) | (0.052) | (0.057) | (0.052) |
| N | 76,452 | | 76,452 | | 76,452 | | 76,452 | |

|  | WRS estimator, exclude Moscow (5) | | WRS estimator, no loan intention (6) | | WRS estimator, adjusted for panel attrition (7) | | |
|---|---|---|---|---|---|---|---|
|  | Formal, t+1 | No job, t+1 | Formal, t+1 | No job, t+1 | Formal, t+1 | No job, t+1 | Survey exit, t+1 |
| Credit market accessibility index, $C_{it}$ | 1.246*** | 1.164*** | 1.291*** | 1.180*** | 1.235*** | 1.128** | 1.139*** |
|  | (0.061) | (0.065) | (0.063) | (0.066) | (0.055) | (0.058) | (0.056) |
| Informal job, t | 0.155*** | 0.703*** | 0.129*** | 0.673*** | 0.143*** | 0.695*** | 0.321*** |
|  | (0.009) | (0.046) | (0.008) | (0.044) | (0.008) | (0.042) | (0.020) |
| No job, t | 0.171*** | 2.916*** | 0.150*** | 2.969*** | 0.160*** | 2.959*** | 0.659*** |
|  | (0.011) | (0.193) | (0.009) | (0.193) | (0.009) | (0.179) | (0.042) |
| (Informal job, t) x $C_{it}$ | 1.217*** | 0.972 | 1.178*** | 0.931 | 1.234*** | 0.981 | 1.023 |
|  | (0.056) | (0.048) | (0.053) | (0.046) | (0.052) | (0.044) | (0.047) |
| (No job, t) x $C_{it}$ | 1.230*** | 1.146*** | 1.201*** | 1.099** | 1.249*** | 1.133*** | 1.027 |
|  | (0.059) | (0.055) | (0.055) | (0.052) | (0.053) | (0.050) | (0.048) |
| N | 71,069 | | 69,280 | | 86,359 | | |

**Notes**: Table reports the *relative risk* of the transition between the employment states from *t* to *t*+1 associated with local credit market accessibility. It shows the results on select variables from seven alternative specifications of the dynamic multinomial logit model of employment. Specification (1) excludes unobserved individual heterogeneity and assumes exogenous initial conditions. Specification (2) includes unobserved heterogeneity but excludes time-averaged variables and initial conditions. Specification (3) uses the Heckman solution to the endogeneity of initial conditions. The full Heckman estimates including equations for initial conditions with exclusion restrictions are shown in Appendix Table A2.1. Specifications (4)-(7) use the WRS solution to the endogeneity of initial conditions and have the same set of covariates as in Table 5. Specification (4) includes additional controls such as real regional GDP per capita (in log), the regional unemployment rate, the 5-year moving average of the regional inflation rate, the log of distance from the community center to the regional capital, and a dummy for whether any state-owned enterprises were closed in the community in the last 12 months. Specification (5) excludes the city of Moscow. Specification (6) re-estimates equation (8) on the sample of individuals who at time *t* did not plan to take a loan in the next 12 months. Specification (7) adds "leaving the survey in *t*+1" as a fourth possible individual outcome. Working informally in *t*+1 is the base outcome in all models. Robust standard errors clustered by individual id are in parentheses; *** $p<0.01$, ** $p<0.05$, * $p<0.1$.

**Table 7. Dynamic Employment Model with Disaggregated Groups of Informal Workers**

|  | WRS estimator | |
|---|---|---|
|  | *Formal job, t+1* | *No job, t+1* |
| Credit market accessibility index, $C_{it}$ | 1.229*** | 1.136** |
|  | (0.062) | (0.064) |
| Employment status, t |  |  |
|   Unregistered employee | 0.196*** | 0.591*** |
|  | (0.016) | (0.059) |
|   Self-employed | 0.111*** | 0.279*** |
|  | (0.015) | (0.050) |
|   Works for a private person | 0.184*** | 0.436*** |
|  | (0.015) | (0.042) |
|   IEA worker | 0.168*** | 1.362*** |
|  | (0.015) | (0.117) |
|   No job | 0.176*** | 2.927*** |
|  | (0.011) | (0.197) |
| (Employment status, t) x $C_{it}$ |  |  |
|   (Unregistered employee, t) x $C_{it}$ | 1.103 | 1.087 |
|  | (0.094) | (0.107) |
|   (Self-employed, t) x $C_{it}$ | 1.117 | 1.372 |
|  | (0.161) | (0.276) |
|   (Works for a private person, t) x $C_{it}$ | 1.178** | 1.100 |
|  | (0.081) | (0.087) |
|   (IEA worker, t) x $C_{it}$ | 1.379*** | 1.164*** |
|  | (0.089) | (0.067) |
|   (No job, t) x $C_{it}$ | 1.234*** | 1.152*** |
|  | (0.059) | (0.054) |

*Post-estimation predictions: Average marginal effect of $C_{it}$ on transition probability*

|  | Employment status in t+1 | | |
|---|---|---|---|
|  | *Formal job* | *Informal job* | *No job* |
| Disaggregated employment status in t |  |  |  |
|   Registered employee | 0.016 | -0.014 | -0.002 |
|   Unregistered employee | 0.031 | -0.034 | 0.002 |
|   Self-employed | 0.027 | -0.049 | 0.022 |
|   Works for a private person | 0.041 | -0.041 | 0.000 |
|   IEA worker | 0.055 | -0.049 | -0.006 |
|   No job | 0.035 | -0.036 | 0.000 |

**Notes**: N=68,057. The sample is restricted to 2006-2014 years Table reports the *relative risk* of the transition between the employment states from *t* to *t*+1 associated with local credit market accessibility. The dependent variable is employment status in *t*+1, with the informal status chosen as the base outcome. The lagged employment status at time *t* is disaggregated by splitting informal workers at time *t* into the four subgroups. In all other ways, the two specifications in this table are the same as in Table 5. Table shows the estimates of equation (8) with correlated random effects and implements the WRS solution to the endogeneity of initial conditions. To save space, we only show the results for the two-way interaction between the disaggregated employment status and the index of credit market accessibility. Robust standard errors clustered by individual id are in parentheses; *** *p*<0.01, ** *p*<0.05, * *p*<0.1.

**Table 8. Predicted Probabilities from Simulated Policies**

|  | Formal, before | Formal, after | Informal, before | Informal, after | Change in informal share |
|---|---|---|---|---|---|
| *Distance to the nearest Sberbank office* |  |  |  |  |  |
| **Policy 1:** open a Sberbank office if no banks within 10 km | 0.630 | 0.649 | 0.142 | 0.132 | -0.010 |
| Starting at 10 km for communities with no banks | (0.016) | (0.020) | (0.012) | (0.015) |  |
| *Distance to the nearest other bank* |  |  |  |  |  |
| **Policy 2:** open a bank office other than Sberbank if no banks within 10 km | 0.626 | 0.655 | 0.160 | 0.121 | -0.039 |
| Starting at 10 km for communities with no other bank | (0.016) | (0.022) | (0.014) | (0.015) |  |
| **Policy 3:** open a second bank office | 0.629 | 0.656 | 0.154 | 0.116 | -0.038 |
| Starting at 10 km for communities with only Sberbank | (0.011) | (0.024) | (0.009) | (0.016) |  |
| *N of bank offices in region* |  |  |  |  |  |
| **Policy 4:** increase to 3 offices per 10,000 for communities with 2+ banks | 0.624 | 0.644 | 0.177 | 0.151 | -0.026 |
| Starting at 2 offices per 10,000 | (0.002) | (0.005) | (0.002) | (0.004) |  |

**Notes**: Table shows the predicted probabilities of being in the formal and informal sectors before and after a given policy is implemented. The predicted probability of being non-employed is $1 - \hat{P}_F - \hat{P}_I$ (not shown). The marginal effects of policies on the employment composition are predicted based on the WRS model estimates shown in Table A2.3. Predicted probabilities are evaluated at the starting values indicated in the table and at sample means of all other covariates. The standard errors are estimated using the delta method and reported in parentheses. The last column shows the change in the relative size of the informal sector due to a simulated policy.

**Table 9. Heterogeneous Effects of Credit Market Accessibility on the Informal Sector**

|  | $d(Informal)/dC$ |  | $Mean(Informal)$ |  |
|---|---|---|---|---|
| Female | -0.018 | (0.003) | 0.122 | (0.003) |
| Male | -0.025 | (0.004) | 0.211 | (0.002) |
| *Real regional average labor earnings, t* |  |  |  |  |
| Lowest tertile | -0.024 | (0.003) | 0.195 | (0.002) |
| Highest tertile | -0.020 | (0.003) | 0.138 | (0.002) |
| *Regional unemployment rate, t* |  |  |  |  |
| Highest tertile | -0.023 | (0.003) | 0.193 | (0.002) |
| Lowest tertile | -0.019 | (0.003) | 0.134 | (0.002) |
| *Credit market accessibility index, t* |  |  |  |  |
| Lowest tertile | -0.024 | (0.003) | 0.195 | (0.003) |
| Highest tertile | -0.021 | (0.003) | 0.151 | (0.002) |
| *Bank availability, t* |  |  |  |  |
| No banks | -0.025 | (0.004) | 0.222 | (0.004) |
| Only Sberbank | -0.021 | (0.003) | 0.163 | (0.004) |
| Other banks | -0.021 | (0.003) | 0.149 | (0.002) |

**Notes**: N=76,452. Based on the WRS estimates shown in Table 5, Table 9 reports the average marginal effect of credit market accessibility on the size of the informal sector, $d(Informal)/dC$, and the mean size of the informal sector by subgroups. Predicted marginal effects are averaged across individuals in a subgroup. The standard errors are estimated using the delta method and reported in parentheses.

**Table 10. Loan Equations with Unobserved Heterogeneity**

|  | Logit | Multinomial logit | |
|---|---|---|---|
|  | Took a loan (t, t+1) | Mortgage and auto loan | Consumer loan |
| Credit market accessibility index, $C_{it}$ | 1.318*** | 1.170*** | 1.366*** |
|  | (0.050) | (0.067) | (0.057) |
| Informal job, $t$ | 0.842*** | 0.739*** | 0.868*** |
|  | (0.039) | (0.058) | (0.044) |
| No job, $t$ | 0.642*** | 0.645*** | 0.629*** |
|  | (0.031) | (0.052) | (0.034) |
| Age, $t$ | 0.962*** | 0.988 | 0.956*** |
|  | (0.012) | (0.020) | (0.013) |
| Age squared, $t$ | 1.023 | 0.979 | 1.034* |
|  | (0.016) | (0.025) | (0.018) |
| Female | 1.128*** | 0.948 | 1.193*** |
|  | (0.043) | (0.054) | (0.051) |
| Russian ethnicity | 1.152** | 0.938 | 1.235*** |
|  | (0.068) | (0.078) | (0.083) |
| Parents' education |  |  |  |
|    Upper vocational | 0.982 | 1.058 | 0.964 |
|  | (0.048) | (0.075) | (0.053) |
|    Higher education | 0.851*** | 0.825** | 0.858** |
|  | (0.047) | (0.067) | (0.053) |
|    Missing | 0.908 | 0.912 | 0.915 |
|  | (0.060) | (0.095) | (0.068) |
| Years of schooling, $t$ | 0.956*** | 1.035*** | 0.930*** |
|  | (0.008) | (0.014) | (0.009) |
| Married, $t$ | 1.125*** | 1.541*** | 1.030 |
|  | (0.044) | (0.096) | (0.045) |
| Number of HH members, $t$ | 1.026 | 1.079*** | 1.012 |
|  | (0.017) | (0.028) | (0.018) |
| Number of kids per HH, 0-13, $t$ | 1.095*** | 0.953 | 1.138*** |
|  | (0.030) | (0.041) | (0.035) |
| Real HH non-durable consumption, log, $t$ | 1.449*** | 1.985*** | 1.322*** |
|  | (0.040) | (0.088) | (0.041) |
| Community population, log | 0.945*** | 0.968* | 0.938*** |
|  | (0.012) | (0.018) | (0.013) |
| Urban | 0.897 | 0.848 | 0.913 |
|  | (0.070) | (0.101) | (0.078) |
| Interval between interviews, $t$ | 0.999 | 0.999 | 1.000 |
|  | (0.001) | (0.001) | (0.001) |
| N | 41,867 | 41,739 |  |

**Notes**: Column 1 presents the relative risk ratios from the random effects logit model for the household's probability of taking a loan between time $t$ and $t+1$. The specification corresponds to equation (9). Columns 2-3 present the relative risk ratios from the multinomial logit model with unobserved heterogeneity. The dependent variable is the type of loan, with "no loan" chosen as the base outcome. Covariates are taken for a randomly chosen household member aged 20-59. The following variables are included but not shown: year dummies, seven federal districts, and the intercept. The omitted categories are "formal sector job" for the employment status and "general secondary education or below" for parents' education. Both models are estimated on the sample of individuals who at time $t$ did not plan to take a loan in the next 12 months. Robust standard errors clustered by household id are in parentheses; *** $p<0.01$, ** $p<0.05$, * $p<0.1$. Variables without subscript $t$ are time-constant.

# Figures

**Figure 1. Timeline**

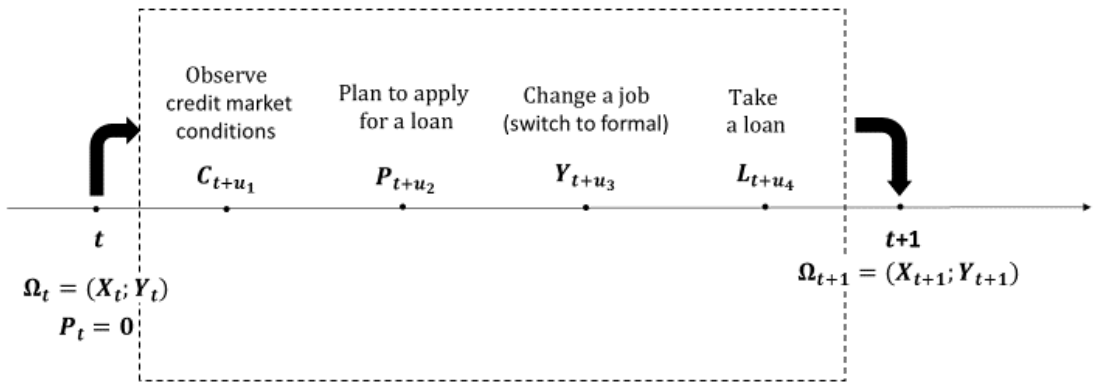

**Figure 2. Share of Loan-Taking Households**

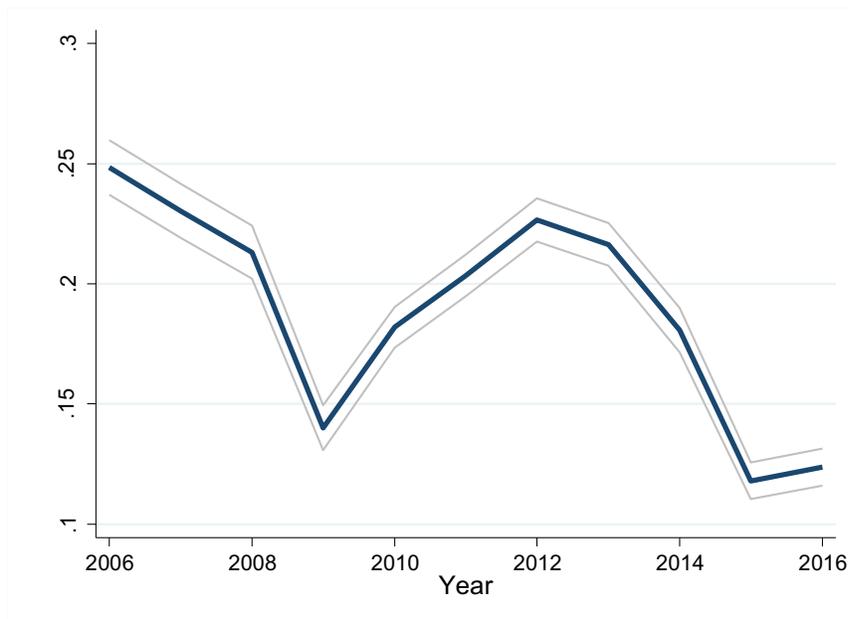

**Notes**: The 95 percent confidence interval is shown in grey.

**Figure 3. Distance to the Nearest Bank Office**

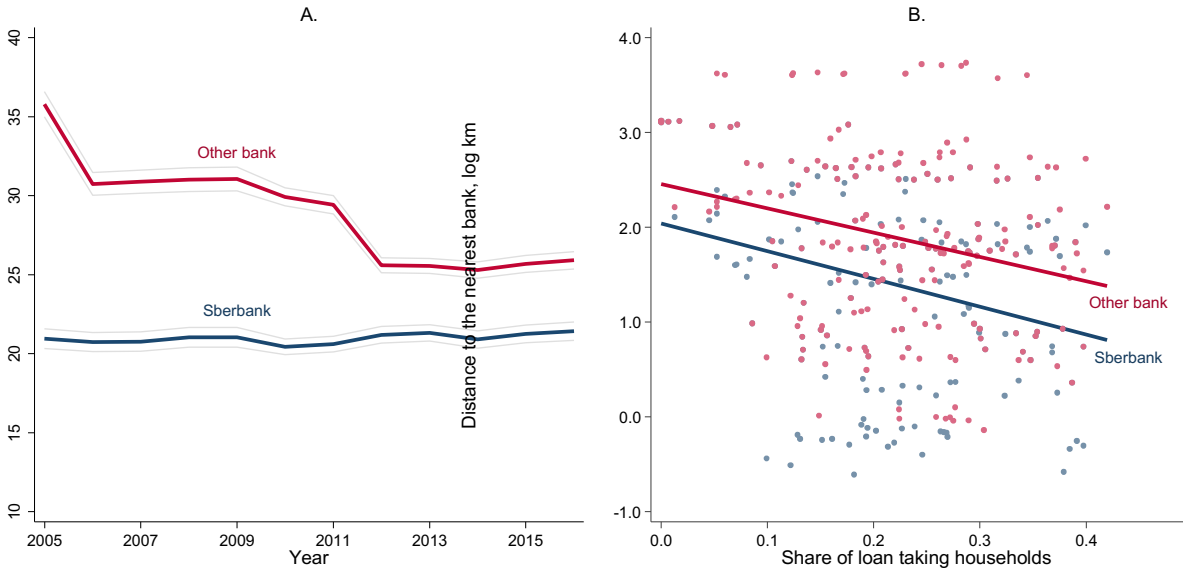

**Notes**: In both panels, the distance to the nearest Sberbank office is depicted in blue, while the distance to the nearest office of other banks is shown in red. The distance measures are conditioned on being non-zero. The sample average with zero values included is presented in Table 3. In Panel A, each distance is averaged by year, and the 95 percent confidence interval is plotted in grey. In Panel B, dots represent region-year observations. Each distance is averaged by region-year. The log scale is used for better data visualization. The variable on the *x*-axis shows the regional share of loan-taking households. The solid line is a fitted line.

**Figure 4. Transition from the Informal to the Formal Sector Before and After Taking a Loan**

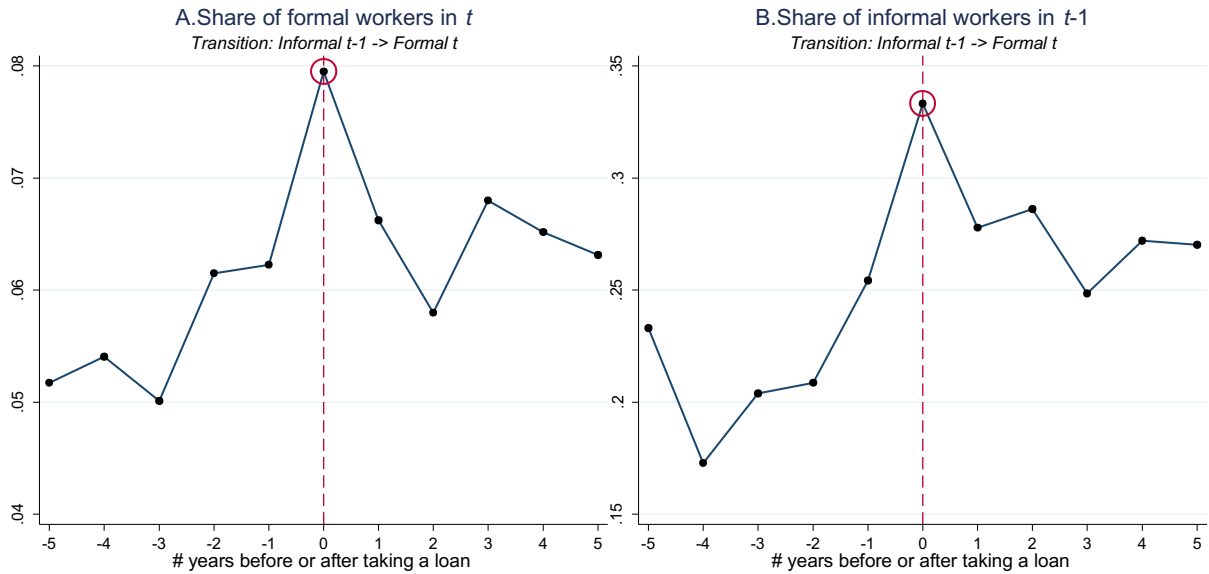

**Notes:** Each dot pertains to the predicted probability of the transition from the informal to the formal sector from an event-study analysis. Panel A depicts the pre dicted rate of entry of informal workers to the formal sector as a share of formal workers in *t*. Panel B plots the predicted switching probability of informal workers in *t*-1 to become formal next year. Event time is defined as year when the household obtains the first loan observed in the data. The probability model is linear and controls for each year before and after the event (the categorical timeline), a quartic polynomial of current age, and calendar year fixed effects. The plotted dots are the coefficients on the timeline variable plus the average predicted share evaluated at timeline=0 (which is the omitted category). The sample is extended to earlier years 2002-2005 to capture as many pre-event points as available, but it is restricted to individuals with non-missing employment status in two consecutive years. Only 5 points before and after the event are shown.

**Figure 5. The Effect of Credit Market Accessibility on Employment Composition**

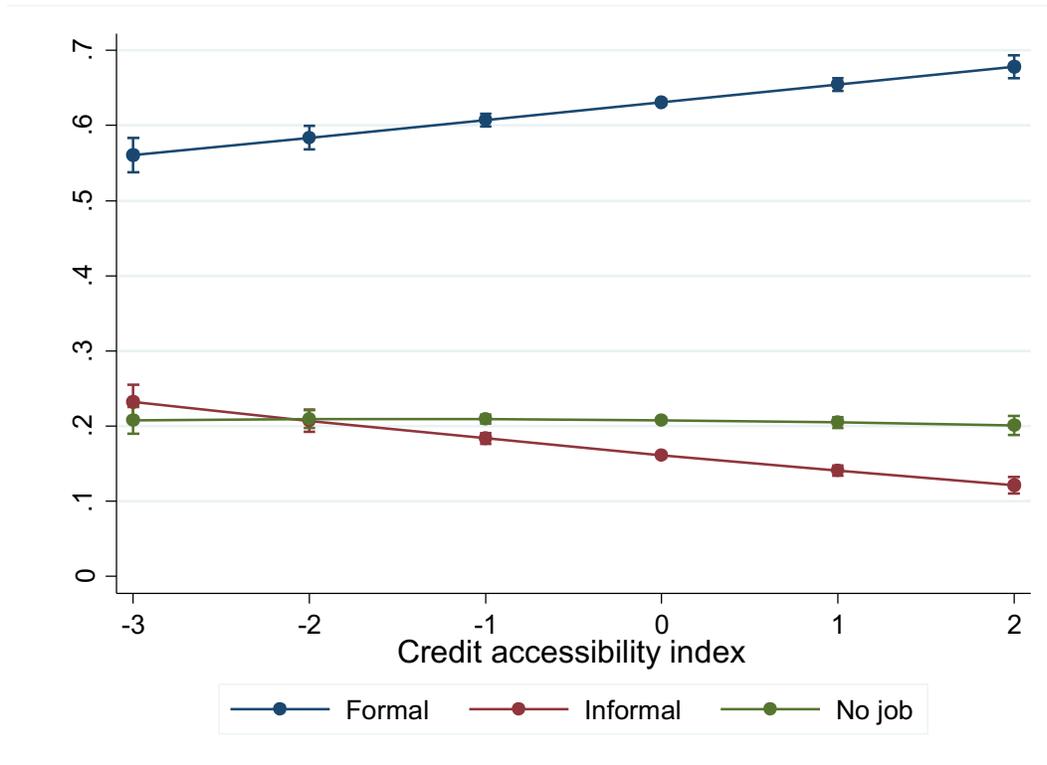

**Notes**: Figure shows the predicted probabilities of being in one of the three employment states at different values of the credit market accessibility index. The index is standardized to have zero mean and a standard deviation of one. Predictions are obtained at sample means of covariates and based on the WRS model reported in Table 5. The standard errors are estimated using the delta method. The 95 percent confidence interval is also shown at each point estimate.

# Appendix 1. Variables

*Employment status*. The three main types of employment status are (1) formal workers, (2) informal workers, and (3) the non-employed. The following flow chart in Figure 6 is created to help the reader to see the sequence of employment questions in the RLMS. The main screening question determines the status of individuals at their primary work. Presently working individuals include those who are either currently working or temporarily absent from work. These employed individuals are further split into those who work at a firm ("an enterprise or organization where more than one person works") and those who work not at a firm.

There are also individuals who answered that they are currently not working but eventually admitted to being engaged in (and paid for) some individual economic activity, such as sewing, taxi driving, babysitting, selling products on the streets, etc. We consider them as part of informal workers.

**Figure 6. The Sequence of Employment Questions in the RLMS**

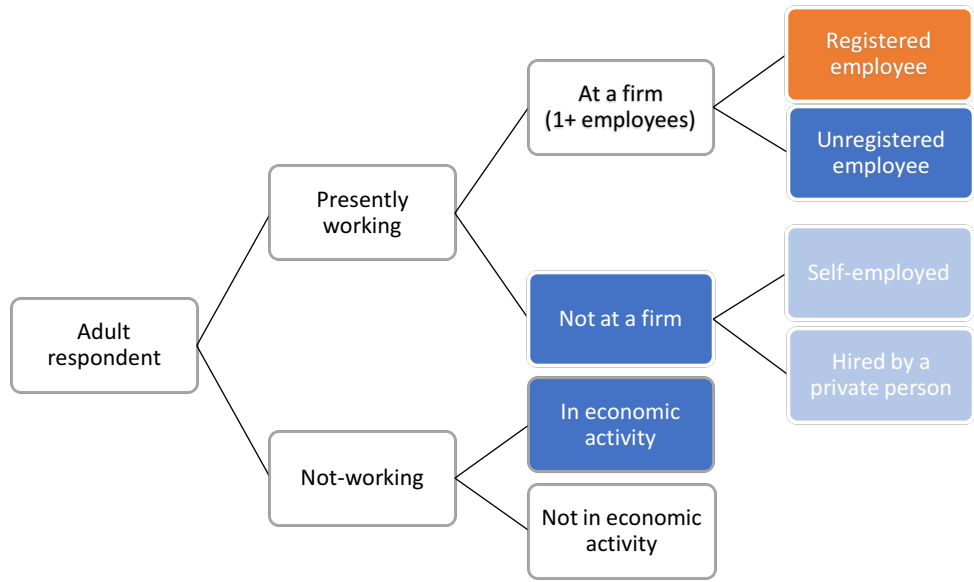

*Formal workers* (in orange) are defined as individuals who are working at a firm with one or more employees and who are officially registered at their primary job. Official registration is determined based on the survey question "Are you on a work roster, written work agreement, or work contract?"

*Informal workers* (in blue) are comprised of (a) unregistered employees, (b) individuals who are not working at an enterprise with more than one person; they could be self-employed or hired by a private person, and (c) individuals who are engaged in individual economic activity (IEA). Among people who are not working at an enterprise with more than one person, the distinction between the self-employed and those hired by a private person can be made in the 2006-2014 surveys.

*Non-employed individuals* include all individuals aged 20-59 without a job. They are not presently working, and they are not engaged in IEA.

*Informal pay.* A categorical variable indicating the share of officially paid earnings: (1) only official pay, (2) partly unofficial pay, and (3) entirely unofficial pay. The variable is created based on the question asked in 2008-2016: "What percent of earnings was received officially, that is, taxes were paid?". This question was answered by individuals who are presently working, worked last month at least one hour and received earnings last month.

*Age, female, year of survey.* Self-explanatory.

*Russian ethnicity.* A binary indicator for Russian ethnicity.

*Parent's education.* A categorical variable indicating the highest level of schooling completed by a parent: [1] "General secondary, lower vocational or less", [2] "Upper vocational", [3] "Higher education or more", and [4] "Unknown level of parents' education". The first category is the base category. The variable is constructed from two sources. The first source is the direct survey question on parents' education asked in 2006 and 2011 waves. The second source is the roster files that match children and parents participating in the RLMS. The roster id allows to determine parents' schooling for children who resided with a parent in the same household in one or more survey waves.

*Years of schooling.* Typical cumulative duration of the highest attained level of education at the time of interview: 0 for no schooling, 4 years for primary general (grades 1-6), 8 years for incomplete secondary (grades 7-9), 9 years for some vocational without a secondary school diploma, 10 years for general secondary (grades 10-11), 11 years for lower vocational with a secondary school diploma, 13 years for upper vocational (technical schools), 15 years for higher education, and 18 years for a post-graduate degree.

*Married.* = 1 for legally married individuals (including those not living together) and 0 for other categories including single, widowed, divorced, and living together without marriage.

*Number of household members.* Counts the number of household members who are presently living in the same household. Calculated from the household roster file.

*Number of kids per household.* Counts the number of children under the age of 14 currently residing in the same household. Calculated from the household roster file.

*Real household non-durable consumption.* The sum of household spending on non-durable goods and services in the last 30 days. The variable is in 2016 prices, in thousand rubles. The construction of this variable follows Gorodnichenko *et al*. (2010).

*Community population.* Number of people living in the community, in thousands. Population in cities is taken from the 2010 Census. Population in villages is taken from the 2010 survey of communities.

*Urban.* =1 for residents of cities or townships (*"the settlement of urban type"*) and 0 for rural residents. The variable is treated as time-constant because the number of urban-rural switchers is very small, 0.3 percent of the estimation sample.

*Interval between interviews.* Number of days between survey interviews. Calculated based on the date of interview.

*Federal districts.* Set of dummies for living in one of the seven federal districts at the time of interview. The districts are Central, Northwest, South, Volga, Urals, Siberia, and Far East.

*First wave fixed effects.* Set of dummies for the starting year of the stochastic sequence or the year of entry to the estimation sample.

*Real regional earnings at age 17.* Average monthly earnings deflated in 2016 prices, in thousand rubles. Available for 1980, 1985, and continuously 1990-2016. The minimum year for age 17 in the estimation sample is 1964. For the 1964-1979 period, regional earnings are imputed by multiplying the national average earnings in each year and the 1980 ratio of regional earnings to the national average, thus assuming the constant regional structure of earnings during the Brezhnev period of the USSR. Missing values in 1981-1984 and 1986-1989 are filled by linear interpolation. Such imputations for the Soviet period may not generate too much noise, as the economy was barely changing and rather stagnant. In addition, we include a binary indicator for the cohorts that reach age 17 before the start of market reforms in 1992.

*Source: Goskomstat Central Statistical Database, 2017. Statistical Yearbook of Russia, Goskomstat (various years).*

*Real regional earnings.* Taken from the same series as above.

*Unemployment rate.* Regional unemployment rate is measured in percent of the labor force according to the standard definition of the International Labor Organization.

*Source: Goskomstat Central Statistical Database, 2017.*

*Regional size of government sector, %.* Percent share of regional employment in state-owned and municipally-owned enterprises.

*Source: Goskomstat Central Statistical Database, 2017.*

*Regional size of informal sector, %.* Percent share of regional employment in the informal sector.

*Source: Обследование населения по проблемам занятости (The Population Survey on Employment Issues), Goskomstat, various years.*

*Any SOEs closed in the community?* A binary indicator for whether any state-owned enterprises were closed in the community during the last 12 months.

*Household took a loan.* =1 if any of household members took a loan in the last 12 months and 0 if no loan is taken.

*Individual plan to obtain a loan.* =1 if the respondent plans to borrow money from a bank in the next 12 months and 0 if the respondent does not have such plans.

*Presence of banks in the community.* This categorical variable takes the value of 1 if no banks, 2 if only Sberbank, and 3 if other banks operate in the community. The third category is the base category. The variable is constructed based on the following two questions from the survey of RLMS communities: "Are there branch offices of any banks, including Sberbank?" and "Are there branch offices of any banks other than Sberbank?"

*Distance to the nearest bank, km.* The distance to the nearest bank is reported for the communities with no banks and for the communities with only Sberbank. If the community has a bank or bank office which does not belong to Sberbank, the distance is coded as zero. To avoid mixing the distances for the first two groups of communities, we create two separate variables by interacting the distance measure with dummies for each group: (1) distance from the community with no banks to the nearest bank and (2) distance from the community with only Sberbank to the nearest other bank. From this module, we use three questions: "Are there offices of any banks, including Sberbank?", "Are there offices of any banks other than Sberbank?", and "How far is it to the nearest bank or bank office, in km?".

*Number of bank offices per 1,000 regional population.* The number of bank offices includes bank headquarters, subsidiary credit organizations, branches, supplementary offices, and operational offices, but excludes cash offices, cash desks, and mobile cash units.

*Source:* Banking Supervision Report, The Central Bank of the Russian Federation, various years*.*

# Appendix 2. Supplementary Tables

## Table A2.1: Dynamic Multinomial Logit Model of Employment Choice

|  | Exogenous initial conditions (1) | | Heckman estimator (2) | | | |
|---|---|---|---|---|---|---|
|  | Formal job, t+1 | No job, t+1 | Formal job, t+1 | No job, t+1 | Formal job, t=1 | No job, t=1 |
|  | 1 | 2 | 3 | 4 | 5 | 6 |
| Credit market accessibility index, $C_{it}$ | 1.182*** | 1.078 | 1.284*** | 1.137** | … | … |
|  | (0.051) | (0.054) | (0.062) | (0.062) | | |
| Informal job, t | 0.055*** | 0.595*** | 0.155*** | 0.694*** | … | … |
|  | (0.004) | (0.039) | (0.009) | (0.042) | | |
| No job, t | 0.078*** | 4.499*** | 0.178*** | 2.775*** | … | … |
|  | (0.004) | (0.255) | (0.010) | (0.170) | | |
| (Informal job, t) x $C_{it}$ | 1.213*** | 0.950 | 1.215*** | 0.966 | … | … |
|  | (0.057) | (0.045) | (0.053) | (0.046) | | |
| (No job, t) x $C_{it}$ | 1.271*** | 1.180*** | 1.243*** | 1.115** | … | … |
|  | (0.057) | (0.054) | (0.057) | (0.052) | | |
| Control variables, $\tilde{X}_i, X_{it}$ | | | | | | |
| Age, t | 0.940*** | 0.753*** | 1.035** | 0.676*** | 1.125*** | 0.618*** |
|  | (0.012) | (0.011) | (0.017) | (0.012) | (0.028) | (0.016) |
| Age squared / 100, t | 1.085*** | 1.503*** | 0.958** | 1.737*** | 0.870*** | 1.891*** |
|  | (0.018) | (0.028) | (0.020) | (0.039) | (0.028) | (0.064) |
| Female | 1.578*** | 2.696*** | 1.568*** | 3.670*** | 1.491*** | 4.081*** |
|  | (0.062) | (0.121) | (0.080) | (0.199) | (0.106) | (0.315) |
| Russian ethnicity | 1.503*** | 1.122* | 1.976*** | 1.065 | 2.265*** | 0.956 |
|  | (0.082) | (0.068) | (0.145) | (0.077) | (0.240) | (0.102) |
| Parents' education | | | | | | |
|    Upper vocational | 1.040 | 1.123** | 1.004 | 1.141* | 0.954 | 1.189* |
|  | (0.051) | (0.064) | (0.065) | (0.077) | (0.090) | (0.118) |
|    Higher education | 1.008 | 1.239*** | 0.952 | 1.348*** | 0.888 | 1.608*** |
|  | (0.057) | (0.079) | (0.070) | (0.104) | (0.093) | (0.176) |
|    Missing | 0.726*** | 0.750*** | 0.834** | 0.840** | 0.861 | 0.881 |
|  | (0.056) | (0.066) | (0.069) | (0.075) | (0.097) | (0.106) |
| Years of schooling, t | 1.170*** | 0.988 | 1.296*** | 0.952*** | 1.365*** | 0.968* |
|  | (0.011) | (0.010) | (0.015) | (0.012) | (0.023) | (0.017) |
| Married, t | 1.357*** | 1.228*** | 1.414*** | 1.191*** | 1.719*** | 1.190** |
|  | (0.052) | (0.054) | (0.068) | (0.060) | (0.128) | (0.093) |
| Number of HH members, t | 0.956*** | 1.058*** | 0.952*** | 1.077*** | 0.931** | 1.114*** |
|  | (0.014) | (0.017) | (0.017) | (0.020) | (0.027) | (0.033) |
| Number of kids per HH, 0-13, t | 1.038 | 0.999 | 1.008 | 1.013 | 0.981 | 1.024 |
|  | (0.027) | (0.029) | (0.032) | (0.034) | (0.053) | (0.058) |
| Real HH non-durable consumption, log, t | 1.023 | 0.854*** | 1.053* | 0.839*** | 1.094* | 0.666*** |
|  | (0.028) | (0.025) | (0.032) | (0.027) | (0.057) | (0.037) |
| Community population, log | 0.942*** | 0.955*** | 0.917*** | 0.943*** | 0.948** | 0.966 |
|  | (0.012) | (0.013) | (0.015) | (0.015) | (0.021) | (0.023) |
| Urban | 0.958 | 0.941 | 1.017 | 0.858 | 1.129 | 0.774* |
|  | (0.075) | (0.082) | (0.102) | (0.089) | (0.155) | (0.109) |
| Interval between interviews, t | 1.001** | 1.000 | 1.001** | 1.000 | | |
|  | (0.001) | (0.001) | (0.001) | (0.001) | | |
| Initial conditions, $R_{i1}$ | | | | | | |
| Relative regional earnings at age 17 | | | | | 1.534** | 1.837*** |
|  | | | | | (0.297) | (0.380) |
| Regional size of government sector, % | | | | | 1.053*** | 1.005 |
|  | | | | | (0.011) | (0.012) |
| Regional size of informal sector, % | | | | | 0.963*** | 0.993 |
|  | | | | | (0.006) | (0.007) |

| | | | |
|---|---|---|---|
| Any SOEs closed in the community? | | 1.001 | 1.165* |
| | | (0.083) | (0.104) |

**Notes**: N=76,452. Table presents the relative risk ratios from the dynamic multinomial logit model of employment with unobserved heterogeneity. The dependent variable is employment status in *t*+1, with the informal status chosen as the base outcome. $C_{it}$ is the aggregate index based on the average of standardized *z*-scores. Columns 1 and 2 report the estimates of equation (8) with correlated random effects, assuming the exogeneity of initial conditions. Columns 3-6 present the results of the Heckman estimator described in Section 3. The following variables are included but not shown: year dummies, seven federal districts, fixed effects for the first year of the stochastic process, and the intercept. The omitted category of parents' education is "general secondary education or below". Robust standard errors clustered by individual id are in parentheses; *** $p<0.01$, ** $p<0.05$, * $p<0.1$. Variables without subscript *t* are time-constant.

**Table A2.2: Dynamic Employment Model Based with the Categories of Informal Pay**

|  | WRS estimator | | |
|---|---|---|---|
|  | **Only official pay, t+1** | **Partly unofficial pay, t+1** | **No job, t+1** |
| Credit market accessibility | 1.370*** | 0.896 | 1.322** |
| index, $C_{it}$ | (0.147) | (0.106) | (0.147) |
| Type of pay, t |  |  |  |
|   Partly unofficial pay | 0.491*** | 2.448*** | 0.964 |
|  | (0.076) | (0.400) | (0.178) |
|   Entirely unofficial pay | 0.052*** | 0.171*** | 0.246*** |
|  | (0.009) | (0.032) | (0.041) |
|   No job | 0.174*** | 0.272*** | 3.919*** |
|  | (0.023) | (0.048) | (0.533) |
| (Type of pay, t) x $C_{it}$ |  |  |  |
|   (Partly unofficial pay, t) x $C_{it}$ | 0.901 | 0.943 | 0.935 |
|  | (0.144) | (0.151) | (0.165) |
|   (Entirely unofficial pay, t) x $C_{it}$ | 1.385** | 1.389* | 1.065 |
|  | (0.199) | (0.236) | (0.146) |
|   (No job, t) x $C_{it}$ | 1.023 | 1.438*** | 0.827** |
|  | (0.099) | (0.197) | (0.080) |

*Post-estimation predictions: Average marginal effect of $C_{it}$ on transition probability*

|  | Payment status in t+1 | | | |
|---|---|---|---|---|
|  | **Only official pay** | **Partly unofficial pay** | **Entirely unofficial pay** | **No job** |
| Payment status in t |  |  |  |  |
|   Only official pay | 0.023 | -0.017 | -0.006 | 0.001 |
|   Partly unofficial pay | 0.033 | -0.036 | -0.003 | 0.007 |
|   Entirely unofficial pay | 0.068 | -0.016 | -0.042 | -0.009 |
|   No job | 0.035 | 0.001 | -0.007 | -0.028 |

**Notes**: N=44,696. The sample is restricted to 2008-2016 years Table reports the *relative risk* of the transition between the payment types from *t* to *t+1* associated with local credit market accessibility. The dependent variable is the payment type in *t+1*, with four categories: (1) only official pay, (2) partly unofficial pay, (3) entirely unofficial pay, and (4) no job. IEA workers and individuals with no labor earnings in the last 30 days are excluded. The third category is chosen as the base outcome. In all other ways, the specification is the same as in Table 5. To save space, we only show the results for the two-way interaction between the payment type and the index of credit market accessibility. Robust standard errors clustered by individual id are in parentheses; *** *p<0.01*, ** *p<0.05*, * *p<0.1*.

**Table A2.3: Different Measures of Credit Market Accessibility**

|  | WRS estimator | |
|---|---|---|
|  | Formal, t+1 | No job, t+1 |
| Type of community, t | | |
|   No banks, t | 0.991 | 1.262 |
|  | (0.325) | (0.499) |
|   Only Sberbank, t | 1.338 | 2.213* |
|  | (0.447) | (0.921) |
| Employment status, t | | |
|   Informal job, t | 0.149*** | 0.934 |
|  | (0.024) | (0.181) |
|   No job, t | 0.128*** | 2.319*** |
|  | (0.023) | (0.427) |
| (Employment status, t) x (Type of community, t) | | |
|   (Informal job, t) x (No banks, t) | 3.436*** | 1.336 |
|  | (1.633) | (0.662) |
|   (No job, t) x (No banks, t) | 2.079 | 1.845 |
|  | (1.051) | (0.944) |
|   (Informal job, t) x (Only Sberbank, t) | 1.842 | 0.668 |
|  | (0.912) | (0.366) |
|   (No job, t) x (Only Sberbank, t) | 1.439 | 0.829 |
|  | (0.776) | (0.442) |
| $C_{it}$ = Distance to the nearest Sberbank office, log km | 1.056 | 1.164 |
|  | (0.096) | (0.132) |
|   (Informal job, t) x $C_{it}$ | 0.752*** | 0.804 |
|  | (0.080) | (0.107) |
|   (No job, t) x $C_{it}$ | 0.779* | 0.747** |
|  | (0.112) | (0.107) |
| $C_{it}$ = Distance to the nearest other bank, log km | 0.872 | 0.789** |
|  | (0.082) | (0.093) |
|   (Informal job, t) x $C_{it}$ | 0.763* | 1.111 |
|  | (0.107) | (0.169) |
|   (No job, t) x $C_{it}$ | 0.872 | 1.012 |
|  | (0.135) | (0.153) |
| $C_{it}$ = N of operating credit organizations in region, log | 20.745*** | 20.283*** |
|  | (12.580) | (14.852) |
|   (Informal job, t) x $C_{it}$ | 1.787 | 0.224* |
|  | (1.247) | (0.192) |
|   (No job, t) x $C_{it}$ | 5.518** | 3.125 |
|  | (4.324) | (2.550) |

**Notes**: N=76,452. Table reports the *relative risk* of the transition between the employment states from *t* to *t*+1 associated with different measures of local credit market accessibility. The specification corresponds to equation (8) with correlated random effects and uses the WRS solution to the endogeneity of initial conditions. The dependent variable is employment status in *t*+1, with the informal status chosen as the base outcome. The only difference from Table 5 is that the aggregate index of credit market accessibility is replaced by the vector of index components. All other covariates are the same as in Table 5. We only show the results for two-way interaction between the lagged employment status and measures of credit market accessibility. Robust standard errors clustered by individual id are in parentheses; *** *p<0.01*, ** *p<0.05*, * *p<0.1*.

**Table A2.4. Loan Equations with Alternative Definitions of the Head of Household**

|  | Took a loan (t, t+1), Logit | | |
|---|---|---|---|
|  | *Random member* | *Oldest male* | *Highest earner* |
| Credit market accessibility index, $C_{it}$ | 1.318*** | 1.296*** | 1.320*** |
|  | (0.050) | (0.053) | (0.051) |
| Informal job, *t* | 0.842*** | 0.863*** | 0.867*** |
|  | (0.039) | (0.040) | (0.039) |
| No job, *t* | 0.642*** | 0.683*** | 0.586*** |
|  | (0.031) | (0.036) | (0.034) |

**Notes**: N=41,867. Table presents the relative risk ratios from the random effects logit model for the household's probability of taking a loan between time *t* and *t*+1. It shows the results on select variables. Each column corresponds to one of the three definitions of the head of household: (1) a random household member, (2) oldest male or oldest female if there are no males in household, and (3) a highest earning member of household. The full model estimates for a random household member are shown in

Table 10. The omitted category is "having a formal sector job" for the employment status. All models are estimated on the sample of individuals who at time *t* did not plan to take a loan in the next 12 months. Robust standard errors clustered by household id are in parentheses; *** $p<0.01$, ** $p<0.05$, * $p<0.1$. Variables without subscript *t* are time-constant.